\documentclass[a4paper,11pt]{article}
\pdfoutput=1 
\usepackage{jcappub} 
\usepackage[T1]{fontenc} 
\usepackage{hyperref}
\usepackage{graphicx}	
\usepackage{amsmath,bm}	
\usepackage{amssymb}	
\usepackage[utf8]{inputenc}
\usepackage{xcolor}
\usepackage{multirow}
\usepackage{soul}
\usepackage[flushleft]{threeparttable}
\usepackage{xspace}
\usepackage{lineno}
\usepackage{url}
\usepackage{rotating}
\usepackage{orcidlink}

\urldef\apodurl\url{https://namaster.readthedocs.io/en/latest/api/pymaster.utils.html#pymaster.utils.mask_apodization}

\def\addtodot#1.#2\relax{#1\rlap{.}^{\dotadd}#2}

\newcommand{\vect}[1]{\ensuremath{\bm{#1}}}
\newcommand{\matr}[1]{\ensuremath{\bm{\mathrm{#1}}}}

\newcommand{\nv}{\hat{\bf n}}

\newcommand{\nmt}{{\tt NaMaster}\xspace}

\title{\boldmath The Simons Observatory: Validation of reconstructed power spectra from simulated filtered maps for the Small Aperture Telescope survey}

\author[a]{Carlos~Herv\'ias-Caimapo\orcidlink{0000-0002-4765-3426},}
\author[b]{Kevin~Wolz\orcidlink{0000-0003-3155-6151},}
\author[b]{Adrien~La~Posta\orcidlink{0000-0002-2613-2445},}
\author[c]{Susanna~Azzoni\orcidlink{0000-0002-8132-4896},}
\author[b]{David~Alonso\orcidlink{0000-0002-4598-9719},}
\author[d,e]{Kam~Arnold\orcidlink{0000-0002-3407-5305},}
\author[f,g,h]{Carlo~Baccigalupi\orcidlink{0000-0002-8211-1630},}
\author[i]{Simon~Biquard\orcidlink{0000-0002-1493-2963},}
\author[j]{Michael~L.~Brown\orcidlink{0000-0002-0370-8077},}
\author[k]{Erminia~Calabrese\orcidlink{0000-0003-0837-0068},}
\author[l,m]{Yuji~Chinone\orcidlink{0000-0002-3266-857X},}
\author[c]{Samuel~Day-Weiss\orcidlink{0009-0003-5814-2087},}
\author[c,n]{Jo~Dunkley\orcidlink{0000-0002-7450-2586},}
\author[a]{Rolando~D\"unner\orcidlink{0000-0003-3892-1860},}
\author[i]{Josquin~Errard\orcidlink{0000-0002-1419-0031},}
\author[k]{Giulio~Fabbian\orcidlink{0000-0002-3255-4695},}
\author[i]{Ken~Ganga\orcidlink{0000-0001-8159-8208},}
\author[k]{Serena~Giardiello\orcidlink{0000-0002-8340-3715},}
\author[o,p]{Emilie~Hertig\orcidlink{0000-0001-9189-4035},}
\author[q]{Kevin~M.~Huffenberger\orcidlink{0000-0001-7109-0099},}
\author[r]{Bradley~R.~Johnson\orcidlink{0000-0002-6898-8938},}
\author[m,s]{Baptiste~Jost\orcidlink{0000-0002-0819-751X},}
\author[t,u]{Reijo~Keskitalo\orcidlink{0000-0001-5748-5182},}
\author[t,u]{Theodore~S.~Kisner\orcidlink{0000-0003-3510-7134},}
\author[v]{Thibaut~Louis\orcidlink{0000-0002-6849-4217},}
\author[w]{Magdy~Morshed\orcidlink{0000-0002-3214-8881},}
\author[c]{Lyman~A.~Page,}
\author[x]{Christian~L.~Reichardt\orcidlink{0000-0003-2226-9169},}
\author[j]{Erik~Rosenberg\orcidlink{0000-0003-3484-5645},}
\author[y]{Max~Silva-Feaver\orcidlink{0000-0001-7480-4341},}
\author[i]{Wuhyun~Sohn\orcidlink{0000-0002-6039-8247},}
\author[c]{Yoshinori~Sueno\orcidlink{0000-0002-3644-2009},}
\author[j]{Dan~B.~Thomas\orcidlink{0000-0003-2244-9530},}
\author[i]{Ema~Tsang~King~Sang\orcidlink{0009-0001-6108-9518},}
\author[i]{Amalia~Villarrubia-Aguilar\orcidlink{0009-0004-4775-9935},}
\author[c,z]{Kyohei~Yamada\orcidlink{0000-0003-0221-2130}}

\affiliation[a]{Instituto de Astrof\'isica and Centro de Astro-Ingenier\'ia, Facultad de F\'isica, Pontificia Universidad Cat\'olica de Chile, Av. Vicu\~na Mackenna 4860, 7820436 Macul, Santiago, Chile}
\affiliation[b]{Department of Physics, University of Oxford, Denys Wilkinson Building, Keble Road, Oxford, OX1 3RH, UK}
\affiliation[c]{Joseph Henry Laboratories of Physics, Jadwin Hall, Princeton University, Princeton, NJ, USA 08544}
\affiliation[d]{Department of Astronomy \& Astrophysics, University of California San Diego}
\affiliation[e]{Department of Physics, University of California San Diego}
\affiliation[f]{The International School for Advanced Studies (SISSA), via Bonomea 265, I-34136 Trieste, Italy}
\affiliation[g]{The National Institute for Nuclear Physics (INFN), via Valerio 2, I-34127, Trieste, Italy}
\affiliation[h]{The Institute for Fundamental Physics of the Universe (IFPU), Via Beirut 2, I-34151, Trieste, Italy}
\affiliation[i]{Universit\'e Paris Cit\'e, CNRS, Astroparticule et Cosmologie, F-75013 Paris, France}
\affiliation[j]{Jodrell Bank Centre for Astrophysics, Department of Physics and Astronomy, University of Manchester, Manchester M13 9PL, UK}
\affiliation[k]{School of Physics and Astronomy, Cardiff University, The Parade,  Cardiff, Wales, UK CF24 3AA}
\affiliation[l]{QUP (WPI), KEK, Tsukuba, Ibaraki 305-0801, Japan}
\affiliation[m]{Kavli Institute for the Physics and Mathematics of the Universe (WPI), UTIAS, The University of Tokyo, Kashiwa, Chiba 277-8583, Japan}
\affiliation[n]{Department of Astrophysical Sciences, Peyton Hall,  Princeton University, Princeton, NJ USA 08544}
\affiliation[o]{Institute of Astronomy, University of Cambridge, Madingley Road, Cambridge, CB3 0HA, United Kingdom} 
\affiliation[p]{Kavli Institute for Cosmology Cambridge, Madingley Road, Cambridge, CB3 0HA, United Kingdom}
\affiliation[q]{Mitchell Institute for Fundamental Physics \& Astronomy and  Department of Physics \& Astronomy, Texas A\&M University, College Station, Texas 77843, USA}
\affiliation[r]{University of Virginia, Department of Astronomy, Charlottesville, VA 22904, USA}
\affiliation[s]{Center for Data-Driven Discovery, Kavli IPMU (WPI), UTIAS, The University of Tokyo, Kashiwa, Chiba 277-8583, Japan}
\affiliation[t]{Lawrence Berkeley National Laboratory, Berkeley, CA, USA}
\affiliation[u]{University of California Berkeley, Berkeley, CA, USA}
\affiliation[v]{Universit\'e Paris-Saclay, CNRS/IN2P3, IJCLab, 91405 Orsay, France}
\affiliation[w]{INFN Sezione di Ferrara, Via Saragat 1, 44122 Ferrara, Italy}
\affiliation[x]{School of Physics, The University of Melbourne, Parkville, VIC 3010, Australia}
\affiliation[y]{Wright Laboratory, Department of Physics, Yale University, New Haven, CT 06511, USA}
\affiliation[z]{Department of Physics, The University of Tokyo, Tokyo 113-0033, Japan}

\emailAdd{carlos.hervias@uc.cl}

\abstract{
We present a transfer function-based method to estimate angular power spectra from filtered maps for cosmic microwave background (CMB) surveys. This is especially relevant for experiments targeting the faint primordial gravitational wave signatures in CMB polarisation at large scales, such as the Simons Observatory (SO) small aperture telescopes. While timestreams can be filtered to mitigate the contamination from low-frequency noise, usual methods that calculate the mode coupling at individual multipoles can be challenging for experiments covering large sky areas or reaching few-arcminute resolution. The method we present here, although approximate, is more practical and faster for larger data volumes. We validate it through the use of simulated observations approximating the first year of SO data, going from half-wave plate-modulated timestreams to maps, and using simulations to estimate the mixing of polarisation modes induced by an example of time-domain filtering. We show its performance through an example null test and with an end-to-end pipeline that performs inference on cosmological parameters, including the tensor-to-scalar ratio $r$. The performance demonstration uses simulated observations at multiple frequency bands. We find that the method can recover unbiased parameters for our simulated noise levels.
}

\begin{document}
\maketitle
\flushbottom

\section{Introduction} \label{sec:introduction}

The cosmic microwave background (CMB) radiation can provide a wealth of information about the very early stages of our Universe. The cosmological community is allocating significant effort to measuring the CMB polarisation with unprecedented sensitivity, as outlined in e.g.,~\cite{2016arXiv161002743A,2019JCAP...02..056A}. Primordial tensor perturbations, in the form of a background of primordial gravitational waves (PGWs) can potentially source a large-scale parity-odd $B$-mode component in the CMB polarisation. In contrast, primordial scalar/density perturbations produce only parity-even $E$-mode fluctuations at linear order (in addition to unpolarised intensity fluctuations)~\cite{1997PhRvL..78.2058K,1997PhRvL..78.2054S}. The amplitude of the power spectrum of the primordial tensor perturbations is parametrised in terms of its ratio with respect to that of scalar perturbations, the so-called tensor-to-scalar ratio $r$. To date, no detection of primordial tensor fluctuations has been made, with the tightest upper bound at 95\% confidence being $r<0.036$ for a pivot scale of $k_p=0.05\,{\rm Mpc}^{-1}$~\cite{2021PhRvL.127o1301A}.

A detection of primordial tensor perturbations, or a sufficiently tight upper bound on $r$, could elucidate many of the remaining questions about the beginning of our Universe, allowing us to discriminate between different models. Cosmic inflation, which posits a short phase of exponentially accelerated expansion at early times, is the most widespread of these models~\cite[e.g.,][]{1980PhLB...91...99S,1982PhRvL..48.1220A,1982PhLB..108..389L}. Within this paradigm, primordial perturbations are seeded by quantum fluctuations in the spacetime metric that are expanded to macroscopic scales during the inflationary period~\cite[e.g.,][]{1981JETPL..33..532M,1982PhRvL..49.1110G,1982PhLB..117..175S,1983PhRvD..28..679B}. The presence of both scalar and tensor fluctuations (i.e. PGWs), is a key prediction of this theory~\cite[e.g.,][]{1984NuPhB.244..541A}. Current and upcoming CMB experiments such as BICEP/KECK~\cite{2021PhRvL.127o1301A,2022ApJ...927...77A}, BICEP Array~\cite{2018SPIE10708E..07H}, SPIDER~\cite{2022ApJ...927..174A}, CLASS~\cite{2014SPIE.9153E..1IE}, {\sc Polarbear}~\cite{2016JLTP..184..805S}, QUBIC~\cite{QUBIC_2022}, GroundBIRD~\cite{2020JLTP..200..384L}, AliCPT-1~\cite{2022JCAP...10..063G}, CMB-S4~\cite{2022ApJ...926...54A}, and LiteBIRD~\cite{2023PTEP.2023d2F01L} are targeting or will target the measurement of these elusive primordial $B$-modes in the next decade.

Polarised foreground emission from our galaxy is a substantial complication for measuring $r$ since it produces a $B$-mode signal much larger than the one from primordial $B$-modes. Thermal dust and synchrotron emission are the two main polarised components that could pose a problem~\cite[e.g.,][]{Planck_2015_X}. Their $B$-mode emission would be equivalent to a tensor-to-scalar ratio much higher than $r\sim0.01$~\cite{Krachmalnicoff_2016}, the target of many upcoming CMB experiments. By exploiting the difference in spectral distributions between the CMB and foregrounds, we can clean the former, using methods collectively known as component separation~\cite{Leach_2008}. However, small residuals left after cleaning may still pose a problem for measuring small values $r \sim 0.001$ in the near future, since the residuals can be as large as the primordial signal we want to measure in the first place. Another relevant foreground contamination are the coherent distortions induced by the deflection of CMB photons passing through the gravitational potential wells of the cosmic large-scale structure, known as CMB weak lensing~\cite{2006PhR...429....1L}. $B$-modes are sourced from primordial CMB $E$-modes at intermediate and small scales~\cite{1998PhRvD..58b3003Z}, effectively creating a lensing $B$-mode spectrum that will dominate over the primordial $B$-modes in all but the largest scales, depending on the value of $r$. For low enough values of $r$, this lensing-induced noise will dominate the sample variance and limit the precision of the cosmological parameters we can measure. We can mitigate this with the technique of delensing, in which we subtract a template of the realization of lensing $B$-modes, reconstructed from an estimator of the lensing potential and observations of the $E$-modes at intermediate and high resolution~\cite{2022PhRvD.105b3511N,2024PhRvD.110d3532H}.

The Simons Observatory (SO) is a ground-based CMB experiment, located at the Cerro Toco site in the Atacama desert in Chile. It targets an assortment of science cases in two main surveys~\cite{2019JCAP...02..056A}. First, a 6-m aperture Large Aperture Telescope (LAT) will focus on small angular-scale CMB science, including high-$\ell$ power spectra and bispectra, secondary anisotropies such as CMB lensing and the Sunyaev-Zel'dovich effect, as well as extragalactic point sources, galaxy clusters, and the time-variable millimetre sky. A second deep survey is being carried out by the Small Aperture Telescopes (SATs), with the main target of measuring or constraining the PGW $B$-mode signal. The SAT observations are covering ${\sim}10$\% of the sky in the same six frequency bands targeted by the LAT, centred at 27, 39\,GHz (LF); 93, 145\,GHz (MF); and 225, 280\,GHz (UHF), aiming to measure the primordial $B$-mode recombination bump ($\ell \sim 80$) from the ground at scales $30 \leq \ell$, constrained by low-frequency noise. The forecasted instrumental sensitivity after 5 years of observations would translate into a measurement of $r$ with $\sigma(r) \simeq 0.002$-$0.003$ after component separation~\cite{2024A&A...686A..16W}.

The nominal SO-SAT experiment comprises 3 small refractive telescopes with a 42\,cm aperture and a field of view of $35^{\circ}$. Each telescope has a 40\,K cryogenic rotating half-wave plate (HWP), 1\,K optics, and a $\sim 100$\,mK focal plane with 7 wafers and ${\sim} 12{,}000$ Transition Edge Sensors~\cite{2024arXiv240505550G}. The nominal instrument configuration will later be expanded to include 3 additional telescopes, potentially using different detector technologies. The UK and Japan are supporting this major enhancement through the SO:UK (two MF SATs) and SO:Japan (one LF SAT) projects, respectively. In this work, we will focus on the two nominal MF bands, 93 and 145\,GHz bands (which we will refer to as f090 and f150, respectively), where most of the sensitivity to the CMB emission is concentrated.

Ground-based observatories contend with multiple sources of low-frequency noise that can affect measurements of the primordial $B$-mode signal. For example, temporal and spatial fluctuations in the emission of the atmosphere due to turbulence and uneven distribution of water vapour, which creates non-trivial correlations between detectors~\cite[e.g.,][]{2015ApJ...809...63E}, as well as by intensity-to-polarisation (I-to-P) leakage due to factors such as thermal drifts, optical mismatch, etc. When doing detector pair differencing with orthogonal antennas, bandpass or beam mismatches arise~\cite[e.g.,][]{2007ApJS..170..263J,2015ApJ...814..110B}. To address these difficulties, the SO-SATs modulate the polarisation signal at frequencies where detector white noise dominates, above atmospheric fluctuations and thermal drifts, using a cryogenic HWP rotating around $f \approx 2$\,Hz~\cite{2014RScI...85c9901K}. This technique also eliminates the need for detector pair differencing, allowing each detector to independently measure the three Stokes parameters $(I,Q,U)$~\cite{2016RScI...87i4503E}. Previous experiments that have used HWP modulation include MAXIPOL~\cite{2007ApJ...665...42J}, ABS~\cite{2018JCAP...09..005K}, and {\sc Polarbear}~\cite{2017JCAP...05..008T}, while CLASS uses a different modulation technology, the variable-delay polarization modulator~\citep[VPM,][]{2025arXiv250111904L}. A demodulation at the modulating fourth harmonics of the HWP rotating frequency, followed by a low-pass filter, is the standard procedure to recover the polarised signal~\cite[e.g.,][]{2024MNRAS.532.2309R}. A further optional high-pass filter can be applied to the polarised timestreams to suppress the remaining low-frequency $1/f$ noise. The impact of all this filtering must be accounted for when generating maps from the filtered timestreams, or the associated bias to the maps must be taken into account and corrected when analysing them~\cite[e.g.,][]{2007ApJ...665...42J,2014ApJ...794..171P,2017ApJ...848..121P,2017A&A...600A..60P,2020ApJ...897...55P}. General filtering to remove low-frequency sources of noise is a standard practice in CMB science, even without polarised modulation, and these effects therefore need to be accounted for in cosmological analyses~\cite[e.g.,][]{2009ApJ...705..978B,2010ApJ...711.1123C,2014ApJ...783...67B,2014PhRvL.112x1101B,2015ApJ...811..126B,2020PhRvD.101l2003S}.

The set of issues described above can be stated generically as the fact that filtered maps are biased due to one or more timestream operations. The objective of this paper is to describe a procedure able to estimate angular power spectra from biased maps from e.g., simulated SO-SAT observations. Specifically, the method described below is based on the estimation of a transfer function describing the loss of power as well as mixing between different polarisation channels due to timestream filtering. This follows similar methods as used for e.g.~ABS~\cite{2018JCAP...09..005K}. We will compare this approach to the exact, but more computationally expensive ``per-$\ell$ method'' that reconstructs the impact of filtering on the full band power window functions. This method, used in past experiments \cite{2014PhRvL.112x1101B,2015ApJ...811..126B}, becomes prohibitively expensive, at least in its brute-force incarnation, for the large data volume that SO will cover. Since the transfer function approach involves non-trivial approximations, in this paper we demonstrate its validity to obtain unbiased power spectra within the uncertainties anticipated for the SO-SAT 1-year deep survey. We also exemplify the limits where the approximations break the method. We note that the pipeline validated here is not the same and uses more aggressive filtering than we anticipate for the analysis of the real SO data.\footnote{Our filtering choices are limited by our software and their implementation of the observation matrix, as will be apparent below. However, we will have no such constraint for the first analysis of real SO data and have much more freedom to implement any required filtering.} The treatment of foreground complexity and component separation is left to future work (see Ref.~\cite{2024A&A...686A..16W} for further details). We validate the methods over realistic simulated timestreams, as opposed to using map-based simulations that a forecast analysis would normally use.

In Section~\ref{sec:methods}, we describe in detail the procedures used to generate filtered maps and their associated unbiased power spectra. In Section~\ref{sec:simulations}, we describe the simulations used to validate our methodology, including the procedure used to simulate SO-SAT observations. Section~\ref{sec:results} presents our results, validating the methods presented in the previous sections and quantifying the residual biases in the measured spectra. In Section~\ref{sec:null_tests}, we demonstrate that this procedure can be used to carry out highly sensitive null tests for the target noise level by the SO-SAT 1-year survey with the target number of detectors. Section~\ref{sec:full_sim} presents an end-to-end analysis from simulated observations to a measurement of $r$ after component separation. We present our conclusions in Section~\ref{sec:conclusions}. All temperatures used in this work are in thermodynamic units relative to the CMB.
\section{Methods} \label{sec:methods}

\subsection{Filter$+$bin mapmaking} \label{sec:filterbin}

Ground-based CMB telescopes typically observe the sky at fixed elevation, scanning back and forth in azimuth, and taking samples at a regular rate with thousands of detectors in the telescope focal plane. While each individual sample is noise-dominated, the sky signal can be teased out by accumulating samples collected over time from the same sky coordinates in a process called ``mapmaking''~\cite[e.g.,][]{1997ApJ...480L..87T}. The relation between a given measured timestream and the true sky is given by
\begin{equation} \label{eq:data_equation}
    \vect{d} = \matr{P} \vect{m} + \vect{n} \text{,}
\end{equation}
where $\vect{d}$ is a vector containing the observed time-ordered samples, $\matr{P}$ is the pointing matrix, which is sparse and relates time samples to coordinates in the sky where the telescope pointed, $\vect{m}$ is the true sky map convolved by the corresponding bandpass and beam, and $\vect{n}$ is the timestream noise.\footnote{This model is simplified as it assumes no difference in beams or bandpasses between detectors. In this study, we will assume that the detector beams and bandpasses are identical and constant in time.} $\matr{P}$ also takes into account the polarisation sensitivity of each detector, i.e. the polarisation angle of the detector projected into sky coordinates, and how much $Q/U$ polarisation a detector measures in a given time sample.

The formal least-square solution to Equation~\eqref{eq:data_equation} is the mapmaking equation, which is linear, minimises the variance, and gives an unbiased\footnote{While this solution is unbiased in the absence of model error bias, real-world experiments are prone to complex effects such as biases from sub-pixel features or mis-modelling of detector gains, among others~\cite{2023OJAp....6E..21N}.} estimation of $\vect{m}$:
\begin{equation}
  (\matr{P}^T \matr{N}^{-1} \matr{P} ) \hat{\vect{m}} = \matr{P}^T \matr{N}^{-1} \vect{d},
\end{equation}
where $\matr{N} = \langle \vect{n} \vect{n}^T \rangle$ is the noise covariance matrix. In most cases, this equation cannot be solved exactly. In practice, the matrix $(\matr{P}^T \matr{N}^{-1} \matr{P} )$ on the left-hand side (LHS) is often impractical to compute, store, or invert explicitly. Firstly, the final matrix is very large:  $(N_{\rm comps} N_{\rm pixels} \times N_{\rm comps} N_{\rm pixels})$, where $N_{\rm pixels}$ is the number of pixels in the map, and $N_{\rm comps}=3$ represents the $I,Q,U$ Stokes components. Secondly, the noise covariance $\matr{N}$ is several orders of magnitude larger, each of its dimensions scaling with the number of detectors and time samples, and exhibiting non-trivial correlations between different detectors and timesteps. To deal with this, optimal mapmaking techniques, like maximum-likelihood (ML) mapmakers, solve the mapmaking equation as a linear system of the form $\matr{A} \vect{x} = \vect{b}$ using iterative methods such as the Preconditioned Conjugate Gradient to invert $\matr{A}$ without explicitly computing or storing it.

This process would be significantly simpler in the presence of a diagonal noise covariance $\matr{N}$, corresponding to the case of white noise and uncorrelated detectors. In this case, 
the LHS matrix is also diagonal across pixels and can be expressed as $N_{\rm pixels}$ ``binned'' matrices of shape $N_{\rm comps}\times N_{\rm comps}$ (or equivalently one $N_{\rm comps}N_{\rm pixels} \times N_{\rm comps}N_{\rm pixels}$ matrix with zero correlation between pixels), which can be easily inverted. The right-hand side (RHS) $\matr{P}^T \matr{N}^{-1} \vect{d}$ is also projected to map space, being a vector with shape $1 \times N_{\rm comps} N_{\rm pixels}$.

Although modulating and demodulating the signal (e.g., using a continuously rotating HWP) should whiten the noise, this idealised situation is never exactly realised. We may nevertheless follow this approach and correct for the impact of the associated approximations a posteriori. In this case, it is common to work with filtered timestreams, in order to deal with systematics and inverse-variance-weight the data. In other words, we apply a set of filters $\matr{F}$ to the data, and project, ``bin'', or ``coadd'' the filtered data $\vect{d}'\equiv\matr{F}\vect{d}$ before solving the mapmaking equation under the assumption of a diagonal noise covariance. The maps resulting from this procedure (commonly called ``filter$+$bin'' mapmaking), and their directly estimated power spectra, will be biased by an unknown multiplicative transfer function (TF). This TF can then be estimated and corrected via simulations, as described in Sections~\ref{sec:per_ell} and \ref{sec:method_tf}. Our goal is to use a relatively fast method that works on the principle of the TF and some approximations.

\subsection{From maps to spectra} \label{sec:maps_to_spectra}
Our objective is to compress the information from maps into efficient summary statistics, such as power spectra $C_\ell$ \cite{1994ApJ...430L..85G,1995PhRvL..74.4369B,1997PhRvD..55.5895T,1998PhRvD..57.2117B,2001PhRvD..64f3001T}. The pseudo-$C_{\ell}$ algorithm (often called the ``MASTER'' algorithm \cite{2002ApJ...567....2H}) is a relatively economic procedure to achieve this goal. We summarize the method here.

We have a full-sky map $\vect{m}(\nv)$ and its spherical harmonic coefficients $m^\alpha_{\ell m}$, where $\alpha$ labels the different polarisation channels ($E$ and $B$). An unbiased and optimal estimator for the power spectra of this map is
\begin{equation}
  \tilde{C}_\ell^{\alpha\beta} \equiv\frac{1}{2\ell+1}\sum_{m=-\ell}^\ell m^\alpha_{\ell m} m^{\beta *}_{\ell m}.
\end{equation}
In order to increase the signal-to-noise of each power spectrum measurement, it is also common to bin the estimated $C_\ell$ into ``band powers'':
\begin{equation}
  \tilde{C}^{\alpha\beta}_b\equiv\sum_{\ell\in b}v_b^\ell \tilde{C}^{\alpha\beta}_\ell,
\end{equation}
where $v_b^\ell$ are the band power weights. Here we use simple top-hat band powers, with $v_b^\ell=\Theta(\ell\in b)/N_b$, where $N_b$ is the number of multipoles in band power $b$, $\Theta$ is a boxcar function, and $\ell \in b$ simply notes the multipole $\ell$ belonging to bin $b$. Note that we distinguish between the true underlying power spectrum $C_\ell^{\alpha\beta}$, the pseudo-$C_\ell$ or ``mode-coupled'' spectrum $\tilde{C}_\ell^{\alpha\beta}$, and the binned version of the latter, $\tilde{C}_b^{\alpha\beta}$.

In the actual data, we have the more complicated case of a masked map $\tilde{\vect{m}}(\nv) \equiv w(\nv)\,\vect{m}(\nv)$, where $w(\nv)$ is a sky mask, which normally will be apodised. The harmonic coefficients of $\tilde{\vect{m}}$ are then a convolution of the true harmonic coefficients $m^\alpha_{\ell m}$ and those of the mask $w_{\ell m}$. Likewise, applying the estimator above to this masked map leads to an estimated power spectrum that is related to that of the true underlying map via a convolution of the form:
\begin{equation} \label{eq:estimated_cl}
  \tilde{C}^{\alpha\beta}_\ell = \sum_{\alpha'\beta'\ell'} W_{\alpha'\beta'\ell'}^{\alpha\beta \ell} C^{\alpha'\beta'}_{\ell'} ,
\end{equation}
where $C^{\alpha'\beta'}_{\ell'}$ is the spectrum of the true map and $W_{\alpha'\beta'\ell'}^{\alpha\beta \ell}$ is the ``mode-coupling matrix'' (MCM). In the case of a simple local operation\footnote{This means that every coordinate in the sky (e.g., a pixel) depends on the local value of a mask and of an unfiltered map in that same coordinate.}, such as masking, the MCM can be calculated analytically as a function of the power spectrum of the mask alone. This forms the basis of the pseudo-$C_\ell$ algorithm. The ``mode-decoupled'' power spectrum, $\hat{C}^{\alpha\beta}_b$, can be computed by inverting the MCM after binning by the band powers $v_b^\ell$:
\begin{equation}
  \hat{C}_b^{\alpha\beta}\equiv\sum_{\alpha'\beta' b'}({\cal W}^{-1})_{\alpha'\beta' b'}^{\alpha\beta b} \tilde{C}^{\alpha'\beta'}_{b'},
\end{equation}
where the binned MCM is
\begin{equation}\label{eq:binned_W}
  {\cal W}_{\alpha'\beta' b'}^{\alpha\beta b}\equiv \sum_{\ell\in b}\sum_{\ell'\in b'}v_b^\ell W_{\alpha'\beta'\ell'}^{\alpha\beta\ell}.
\end{equation}
Thus, the relation between the mode-decoupled spectrum and the true underlying spectrum is:
\begin{equation}\label{eq:cl_bbl}
  \hat{C}_b^{\alpha\beta}\equiv\sum_{\alpha'\beta'\ell'}B_{\alpha'\beta' \ell'}^{\alpha\beta b} C^{\alpha'\beta'}_{\ell'},
\end{equation}
where the ``band power window functions'' (BPWFs) \cite{1999PhRvD..60j3516K} are given by:
\begin{equation}\label{eq:bbl_mask}
  B_{\alpha'\beta' \ell'}^{\alpha\beta b}\equiv \sum_{\alpha''\beta'' b''}({\cal W}^{-1})_{\alpha''\beta'' b''}^{\alpha\beta b}\sum_{\ell''\in b''}v_{b''}^{\ell''}\,W_{\alpha'\beta' \ell'}^{\alpha''\beta'' \ell''}.
\end{equation} 

Finally, the case of a filtered and masked map is given by
\begin{equation} \label{eq:obs_matrix}
  \tilde{\vect{m}}(\nv)\equiv w(\nv)\,[\matr{R}\,\vect{m}](\nv) \text{,}
\end{equation}
where $\matr{R}$ is the filtering operator, often called the ``observation matrix'' \cite{2014PhRvL.112x1101B}, relating the observed and true skies. In our case, $\matr{R}$ results from the projection and binning of all timestream filtering operations onto the sky. This observation matrix will induce additional effects, suppressing and mixing polarisation modes in a non-trivial way beyond the effect of masking. As before, the power spectrum of $\tilde{\vect{m}}$ is related to the true spectrum via a mode-coupling matrix that we will label $X_{\alpha'\beta'\ell'}^{\alpha\beta\ell}$ to distinguish it from the mask-only MCM $W_{\alpha'\beta'\ell'}^{\alpha\beta\ell}$. Unlike in the simple case of only masking, the MCM for a general filter$+$mask operator cannot be calculated analytically. We may, however, calculate the power spectrum of $\tilde{\vect{m}}$ by ignoring the presence of $\matr{R}$ and considering only the impact of the mask. The resulting estimator will still relate to the true power spectrum through a set of BPWFs, as in Equation~\eqref{eq:cl_bbl}, although the BPWFs are given by
\begin{equation}\label{eq:bbl_obs}
  B^{\alpha\beta b}_{\alpha'\beta' \ell'} \equiv \sum_{\alpha''\beta'' b''}({\cal X}^{-1})_{\alpha''\beta'' b''}^{\alpha\beta b}\sum_{\ell''\in b''}v_{b''}^{\ell''}\,X_{\alpha'\beta' \ell'}^{\alpha''\beta'' \ell''},
\end{equation}
as opposed to Equation~\eqref{eq:bbl_mask}. Here ${\cal X}$ is the binned version of $X$, in analogy to ${\cal W}$ in Equation~\eqref{eq:binned_W}. Since we cannot calculate $X$ analytically, we must resort to a simulation-based approach in order to estimate the BPWFs. The next two sections present two different approaches to achieve this. 

\subsubsection{Mode coupling per-$\ell$} \label{sec:per_ell}

Following Ref.~\cite{2014PhRvL.112x1101B}, the BPWFs can be reconstructed from the response of each band power to a signal map exciting a single multipole $\ell$. See also Ref.~\cite{2022ApJ...928..109L}. In more detail:
\begin{enumerate}
    \item We generate a scalar full-sky simulation from a unit delta-like power spectrum at multipole $\ell_*$ (i.e. $C_{\ell_*}=\delta_{\ell \ell_*}$).
    \item By assigning the harmonic coefficients of the above delta-like scalar simulation to either the $E$- or the $B$-mode component of a polarised map with only zeros otherwise, we effectively generate pure-$E$ and pure-$B$ simulations.
    \item We ``observe'' the resulting maps, applying to them the same processing and filters used on the real data, and masking them.
    \item We estimate the band powers of the resulting maps following the steps described in the previous section. Specifically, let $C^{(\alpha_*\beta_*),\alpha\beta}_{(\ell_*),b}$ be the $b$-th band power of the cross-spectrum between the $\alpha$ and $\beta$ polarisation channels resulting from cross-correlating the pure-$\alpha_*$ and pure-$\beta_*$ observed $\delta$-like maps produced to excite the multipole $\ell_*$.
    \item From Equation~\eqref{eq:cl_bbl}, it is clear that $C^{(\alpha_*\beta_*),\alpha\beta}_{(\ell_*),b}$ is an estimator for the BPWF $B_{\alpha_*\beta_*\ell_*}^{\alpha\beta b}$.
\end{enumerate}
We repeat this procedure over simulations and average over them to increase the accuracy of the estimated BPWFs. In order to reduce the number of simulations needed to achieve a given precision in the estimated BPWFs, we generate them from distributions with a smaller 4-point function rather than a Gaussian (since the BPWFs are estimated from power spectra). Specifically, we generate them from a Bernoulli distribution (i.e. the real and imaginary elements of the resulting harmonic coefficients for $\ell=\ell_*$ are $\pm 1/\sqrt{2}$ with equal probability), which achieves the lowest possible kurtosis (and hence the lowest variance for any quantity that involves quadratic combinations of these maps) for maps with independent pixel values~\cite{Hutchinson01011989}.

\begin{figure}
    \centering
    \includegraphics[width=1.0\textwidth]{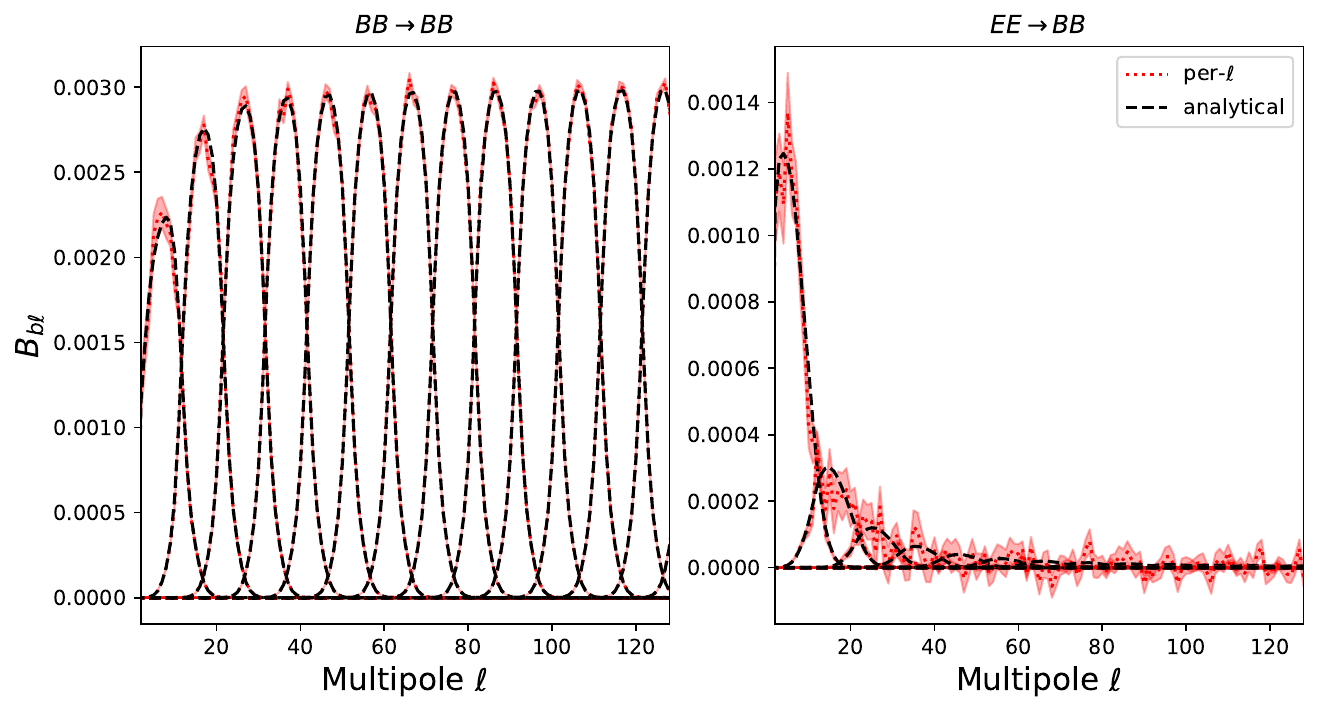}
    \caption{Comparison of the reconstructed band power window functions using the per-$\ell$ simulation-based method described in Section~\ref{sec:per_ell}. The dashed black line shows the analytical \nmt-calculated BPWF for $BB \rightarrow BB$ (left hand side panel) and $EE \rightarrow BB$ (right-hand side panel) for a partial sky mask at $N_{\rm side}=64$. The dotted red line shows the simulation-based BPWFs averaged across 100 simulations per $\ell$. The red shaded area represents the standard deviation across the 100 simulations divided by $\sqrt{100}$. For this example, we use the mask constructed from the hits map in Figure~\ref{fig:hits_full}. The other BPWFs look qualitatively similar. We remind the reader this is an example of a test case with no time-domain filtering.
    }
    \label{fig:compare_mcm}
\end{figure}

We can partially validate this method in a simple way, by considering the limit in which no filtering operations are applied to the data beyond masking. In this case, the MCM and the BPWFs can be estimated analytically using the standard pseudo-$C_\ell$ algorithm via \nmt~\cite{2019MNRAS.484.4127A}. We can then compare this analytical solution with the estimate from $\delta$-like simulations. Figure~\ref{fig:compare_mcm} shows the $BB \rightarrow BB$ and $EE \rightarrow BB$ BPWFs estimated through these two methods, with the analytical solution as the dashed black line, and the simulation-based estimate as the dotted red line, using 100 simulations per $\ell$. Although the simulation-based estimate is intrinsically noisy, the differences between them are consistent within the sample variance due to the limited number of simulations.

The main drawback of this method is the large number of simulations needed to reconstruct the BPWF. As we will show in Section~\ref{sec:bpwf}, an accurate estimate of the BPWFs, particularly accounting for $E/B$ leakage, may require $\mathcal{O}(10^5)$ simulations, on which expensive filtering operations must be carried out, in order to reconstruct the first ${\sim} 200$ multipoles for SO-SAT. This is almost impossible to achieve when doing end-to-end simulated filtering\footnote{As an illustration, it takes approximately 20 minutes to filter and bin one map as described in Section~\ref{sec:using_toast} with one node of the \textsc{perlmutter} computer at NERSC, using 128 cores. To reconstruct the first 200 multipoles, we would need to repeat this for $191\times4\times1000=764,000$ maps, assuming we do 1000 realizations per-$\ell$. Often, we will need even more realizations to constrain the leakage terms. In this example, this would take $10,611$ node-days.}. This process may be accelerated through the use of observation matrices (e.g., as done by the BICEP collaboration~\cite{2014PhRvL.112x1101B}), as long as constructing this matrix is feasible given available computing resources. This may pose challenges for experiments covering large sky areas, such as SO-SAT, or reaching few-arcminute resolutions. This motivates exploring alternative approximate but less computationally demanding methods.

\subsubsection{Using a transfer function} \label{sec:method_tf}
General filtering operations can lead to both mode mixing and the statistical coupling of $E$- and $B$-modes ($E/B$ leakage). However, if we assume that mode coupling contributions are dominated by the sky mask, while mode suppression (no mixing) and leakage are driven by filtering, the number of simulations required to estimate the BPWFs is significantly reduced. None of the filtering operations studied here, at least in their idealised implementations, would lead to I-to-P leakage.

Specifically, we will assume that the binned pseudo-$C_\ell$s are related to the true $C_\ell$s through a combination of a mode-coupling matrix due to the mask, and a TF caused by filtering. Thus, we approximate the true mode-coupling matrix for the pseudo band powers in equation~\eqref{eq:bbl_obs} as
\begin{equation} \label{eq:mcm_tf}
 \sum_{\ell\in b}v_{b}^{\ell}\,X_{\alpha'\beta' \ell'}^{\alpha\beta \ell} \approx \sum_{\gamma\delta}T^{(\alpha\beta)(\gamma\delta)}_b\sum_{\ell\in b}v_{b}^{\ell}\,W_{\alpha'\beta' \ell'}^{\gamma\delta \ell}
\end{equation}
where $W_{\alpha'\beta' \ell'}^{\gamma\delta \ell}$ is the mode-coupling matrix due solely to the mask (and estimated quickly with \nmt), and $T^{(\alpha\beta)(\gamma\delta)}_b$ is the TF. This equation implies that the masking happens first, while the filtering happens after. Of course, in a real survey, this happens in the opposite order. This simplification is done to estimate a TF depending on a reduced number of band power bins rather than a much larger number of individual multipoles. In Section~\ref{sec:bpwf}, we validate this approximation.

 The main assumption here (that coupling between different harmonic modes is driven by the sky mask, and not by filtering) is not entirely without basis. Consider the toy scenario of a one-dimensional sky mapped by a continuously scanning telescope. Filtering the resulting timestream is, in the most common cases (e.g. Fourier-space filters, polynomial filters), a translationally-invariant linear transformation. As such, it affects each Fourier mode of the timestream (and of the 1-dimensional sky) independently, without mixing them. This is similar to beam smoothing, although most filters instead affect the largest scales. This analogy breaks down for the true two-dimensional sky, especially in regions of poor cross-linking, and when filtering occurs over finite time periods (e.g. telescope subscans), breaking translational invariance. Nevertheless, it illustrates the reasons for the good performance of the approximate transfer function approach we will demonstrate in Section~\ref{sec:results}.

In the case of scalar fields, the TF has a very simple interpretation: it quantifies the level to which filtering suppresses the power of different angular scales and can be estimated from the ratio of the binned power spectra of filtered and unfiltered simulations. In the case of spin-$2$ polarised maps, we must also take into account the leakage between polarisation channels caused by filtering. The transfer function is a 4$\times$4 matrix for each band power $b$, with estimates computed as follows:

\begin{enumerate}
    \item We generate full-sky Gaussian realisations of a known input power spectrum, which can be anything in principle. In our case, we use a power-law (PL) spectrum with index $-2$, which we denote $C_{\ell}^{\rm PL}$ (specifically, $C_\ell^{\rm PL}\equiv(\ell+10)^{-2}$). Having generated a set of harmonic coefficients from this power spectrum, we create ``pure-$E$'' and ``pure-$B$'' spin-2 simulations by assigning these coefficients to either the $E$- or the $B$-mode component of a polarised full-sky field, respectively.
    
    \item We filter and mask these simulations as described in the previous section for the $\delta$-like simulations.
    
    \item We then compute the angular power spectra of the filtered and masked simulations, and those of the same simulations with only the mask applied. For clarity, $\tilde{C}^{{\rm PLF},(\alpha\beta)(\gamma\delta)}_b$ denotes the binned pseudo-$C_\ell$ estimate of the pure-$\alpha$ and pure-$\beta$ filtered-and-masked simulations for the pair of polarisation channels $(\gamma,\delta)$. Likewise, $\tilde{C}^{{\rm PL},(\alpha\beta)(\gamma\delta)}_b$ denotes the same binned pseudo-$C_\ell$ for the unfiltered (but masked) simulations, which can be calculated analytically with \nmt.
    
    \item We average both sets of power spectra over a number of simulations ($100$ in our case). Note that $\tilde{C}^{{\rm PL},(\alpha\beta)(\gamma\delta)}_b$ does not need to be calculated empirically averaging over realisations, but instead can be alternatively calculated analytically through multiplying the input spectrum by the mask MCM, as illustrated by Equation~\eqref{eq:estimated_cl}.
\end{enumerate}

As discussed above, the power spectra for the filtered and unfiltered simulations are related to one another by the TF:
\begin{equation}
\tilde{C}^{{\rm PLF},(\alpha\beta)(\gamma\delta)}_b=\sum_{\alpha'\beta'}T_b^{(\alpha\beta)(\alpha'\beta')}\tilde{C}^{{\rm PL},(\alpha'\beta')(\gamma\delta)}_b.
\end{equation}
For a fixed combination of $(\alpha,\beta)$, the equation above may be expressed as a matrix-vector multiplication:
\begin{equation} \label{eq:TF_linear}
\vect{b}^{(\alpha \beta)}_b = \matr{C}_b \vect{t}^{(\alpha \beta)}_b \text{,}
\end{equation}
where the vectors $\vect{b}$ and $\vect{t}$ consist of the power spectra of filtered simulations and the TF respectively,
\begin{align}
&\vect{b}^{(\alpha \beta)} = \left( \tilde{C}_b^{{\rm PLF}, (\alpha \beta)(EE)}, \tilde{C}_b^{{\rm PLF}, (\alpha \beta)(EB)}, \tilde{C}_b^{{\rm PLF}, (\alpha \beta)(BE)}, \tilde{C}_b^{{\rm PLF}, (\alpha \beta)(BB)}  \right) \text{.}\\
&\vect{t}^{(\alpha \beta)} = \left( T_b^{(\alpha \beta)(EE)}, T_b^{(\alpha \beta)(EB)}, T_b^{(\alpha \beta)(BE)}, T_b^{(\alpha \beta)(BB)}  \right),
\end{align}
and the matrix $\matr{C}_b$ is the collection of unfiltered binned pseudo-$C_\ell$s
\begin{equation}
(\matr{C}_b)_{(\gamma \delta)(\alpha \beta)} = \tilde{C}_b^{{\rm PL}, (\alpha \beta)(\gamma \delta)} \text{.}
\end{equation}
Since the matrix $\matr{C}$ is not necessarily invertible, we estimate the TF by solving Equation~\eqref{eq:TF_linear} via least-squares fitting
\begin{equation} \label{eq:tf_least_squares}
\hat{\vect{t}}_b^{(\alpha \beta)} = \left[ \matr{C}^T_b \matr{C}_b \right]^{-1} \matr{C}_b^T \cdot \vect{b}^{(\alpha \beta)}_b \text{.}
\end{equation}

With the TF estimated as we just described, we can construct our model for the mode-coupling matrix of the band powers by combining it with the mask-only mode-coupling matrix (Equation~\ref{eq:mcm_tf}). The method then proceeds as it does for the standard pseudo-$C_\ell$ approach: we bin and invert this mode-coupling matrix, and use it to estimate the mode-decoupled band powers and the associated BPWFs. Our implementation of this method\footnote{\url{https://github.com/simonsobs/SOOPERCOOL} v0.0} includes all the cross-spectra of the $IEB$ fields. As stated, this paper will only focus on polarisation and ignore intensity.
\section{Simulations} \label{sec:simulations}

In this section, we describe the simulations used to estimate the transfer function (TF), as well as validate the methods described. We describe the input maps we use, the scanning strategy of our simulated SO-SAT deep survey, and how we produce filter+bin maps.

\subsection{Input simulated maps} \label{sec:input_simulations}

All of our simulated maps are in \textsc{healpix} pixels~\cite{2005ApJ...622..759G}, produced at single float precision. For a comparison between the numerical per-$\ell$ and TF methods, shown in Section~\ref{sec:bpwf}, we use maps at resolution $N_{\rm side}=128$. The limited resolution is necessary in order to compute and store the $3N_{\rm pixels} \times 3N_{\rm pixels}$ observation matrix used to incorporate the effects of filtering (see Section~\ref{sec:obsmat}). Handling this matrix rapidly becomes challenging for resolutions $N_{\rm side}\geq 256$. For the TF method results at full resolution and multipole range, shown in Section~\ref{sec:tf_full_multipole}, we use maps at resolution $N_{\rm side}=512$. All the angular power spectra in this work are calculated with \nmt, which corrects for the mode coupling arising from partial sky observation. We do not use $E$/$B$-mode purification for simplicity. Incorporating the standard $B$-mode purification techniques that correct the $E$-to-$B$ leakage caused by the sky mask \cite{2006PhRvD..74h3002S} is trivial to implement in this formalism (one simply replaces all fields and power spectra by their purified versions). Filtering leads to additional leakage that, ideally, should also be corrected at the map level. Methods for this exist, exploiting observation matrices to construct the covariance of the filtered data \cite{astro-ph/0207338,2014PhRvL.112x1101B}. This is an area of active research in SO, where the larger sky area covered makes deploying these techniques challenging. The masks are constructed from normalised hits maps and with a C1 apodisation~\cite{2009PhRvD..79l3515G} with a $5^{\circ}$ length.\footnote{See \apodurl~for details.}

For the per-$\ell$ method, we generate 4 maps (one per each polarisation $EE, EB, BE, BB$) of correlated pure modes that excite one $\ell$ at a time, per each realisation, per each $\ell$. We filter them using the $N_{\rm side}=128$ observation matrix of our observation schedule, which are described in Section~\ref{sec:obsmat}. We create 6144 realisations per $\ell$.

For the TF method, all the maps we create have the corresponding pixel window function and a Gaussian beam smoothing with full width at half maximum (FWHM) of $30^\prime$ for the f090 band and $17^\prime$ for the f150 band. The power spectrum of the transfer function simulations, in general, should have the shape of the power spectrum of the unfiltered real sky signal. We estimate the TF from 100$\times$2 input maps generated with a power-law theory spectrum $C_{\ell} = (\ell+10)^{-2}$, which is similar to the angular power spectrum of galactic foregrounds (which are non-Gaussian also). Each realisation has pure-$E$ modes with $a_{\ell m}^B=0$ or pure-$B$ modes with $a_{\ell m}^E=0$. We also produce validation simulations. We use the estimated TF to reconstruct the polarisation spectra of these validations and check the consistency with input theory spectra. We simulate 100 CMB maps with a fiducial spectrum consistent with the \textit{Planck} 2018 best-fit parameters \cite{2020A&A...641A...6P} and lensing as the only source of $B$-modes (therefore with tensor-to-scalar ratio $r=0$). To assess the limitations of the TF method, we also produce another set of 100 validation simulations with a steeper power-law with theory spectrum $C_{\ell} = (\ell+10)^{-6}$.

\subsubsection{Foreground simulations}

In Sections~\ref{sec:null_test_signal_noise} and \ref{sec:full_sim}, we consider foregrounds and component separation. For that, we use the \textit{Planck} 353\,GHz channel as ancillary data. We use the same CMB realisation with lensing-only $B$-modes and smoothed to a $5^\prime$ beam. We include two noise maps per realisation, corresponding to the detector A and B focal plane splits using the end-to-end \textsc{npipe} noise simulations~\cite{NPIPE2020}.\footnote{Available at NERSC at \path{/global/cfs/cdirs/cmb/data/planck2020/npipe}.} We observe and filter 353\,GHz CMB$+$noise maps. By this, we mean transforming the map into a timestream, filtering it with the same SO-SAT filters, and then binning the map as described in Section~\ref{sec:filterbin}. We construct 100 \textit{Planck}-like 353\,GHz maps.

We use full-sky templates for synchrotron and thermal dust from \textsc{pysm}~\cite{2017MNRAS.469.2821T}. We use the \texttt{s1} model for synchrotron and the \texttt{d1} model for dust, which include spatially anisotropic foreground amplitudes and, in addition, moderate spatial variation in the frequency spectral energy distributions (SED). Real observations inform these models and therefore are available as single realizations. For covariance estimation purposes, we also create 100 realisations of polarised dust emission as a random Gaussian field, with a power-law spectrum at 353\,GHz given by
\begin{equation} \label{eq:dust_Dell}
    \mathcal{D}_{\ell}^{BB} = A_{\rm d}^{BB} (\ell / 80)^{\alpha_{\rm d}^{BB}} \text{,}
\end{equation}
where $A_{\rm d}^{BB} = 35$\,$\mu$K$^2$ and $\alpha_{\rm d}^{BB}=-0.16$ \cite{2024A&A...686A..16W}\footnote{While this reference used $A_{\rm d}^{BB} = 28$\,$\mu$K$^2$, 35\,$\mu$K$^2$ is a better fit to the mask used in this work.}. The dust maps at 353\,GHz are then scaled to other frequencies using a modified black body (MBB) SED with spatially constant spectral index $\beta_{\rm} = 1.54$ and temperature $T_{\rm d}=20$\,K.

\subsection{Simulated dataset} \label{sec:schedule}

We simulate an approximate 1-year SO-SAT deep survey. To reduce the computational cost, we simulate the observation of the South Galactic patch performed by a single MF SAT on the first day of each month during a year. We assume no observations are made in the time block from the 15\textsuperscript{th} of January to the 15\textsuperscript{th} of March, since this accounts for possible bad weather conditions and downtime for hardware upgrades. Thus, in practice, the observation scanning strategy simulated here includes 10 days of observations. We chose this simplified schedule, containing a fraction of the real schedule, to reduce the computational requirements of the simulation while evenly sampling the temporal variations in the scanning pattern due to sun avoidance across the year.

As an example scanning strategy, we observe at three different fixed elevations of $50, 55, 60^{\circ}$. The South patch is prioritised, so it is observed from the moment it rises above the limiting elevation. The real SAT scan pattern will also include observations of the North Galactic patch when the South patch is not visible, so our simulated strategy is not optimised while being fairly representative of what our real observations will look like. An individual scan is 50 minutes long on average, covering a typical azimuth range of $40^{\circ}$. The azimuth scan speed is 1\,deg\,s$^{-1}$, while the scan acceleration is 1\,deg\,s$^{-2}$. We warn the reader this describes an example scanning strategy and may not represent how SO actually scans in real observations.

\begin{figure}
    \centering
    \includegraphics[width=0.9\textwidth]{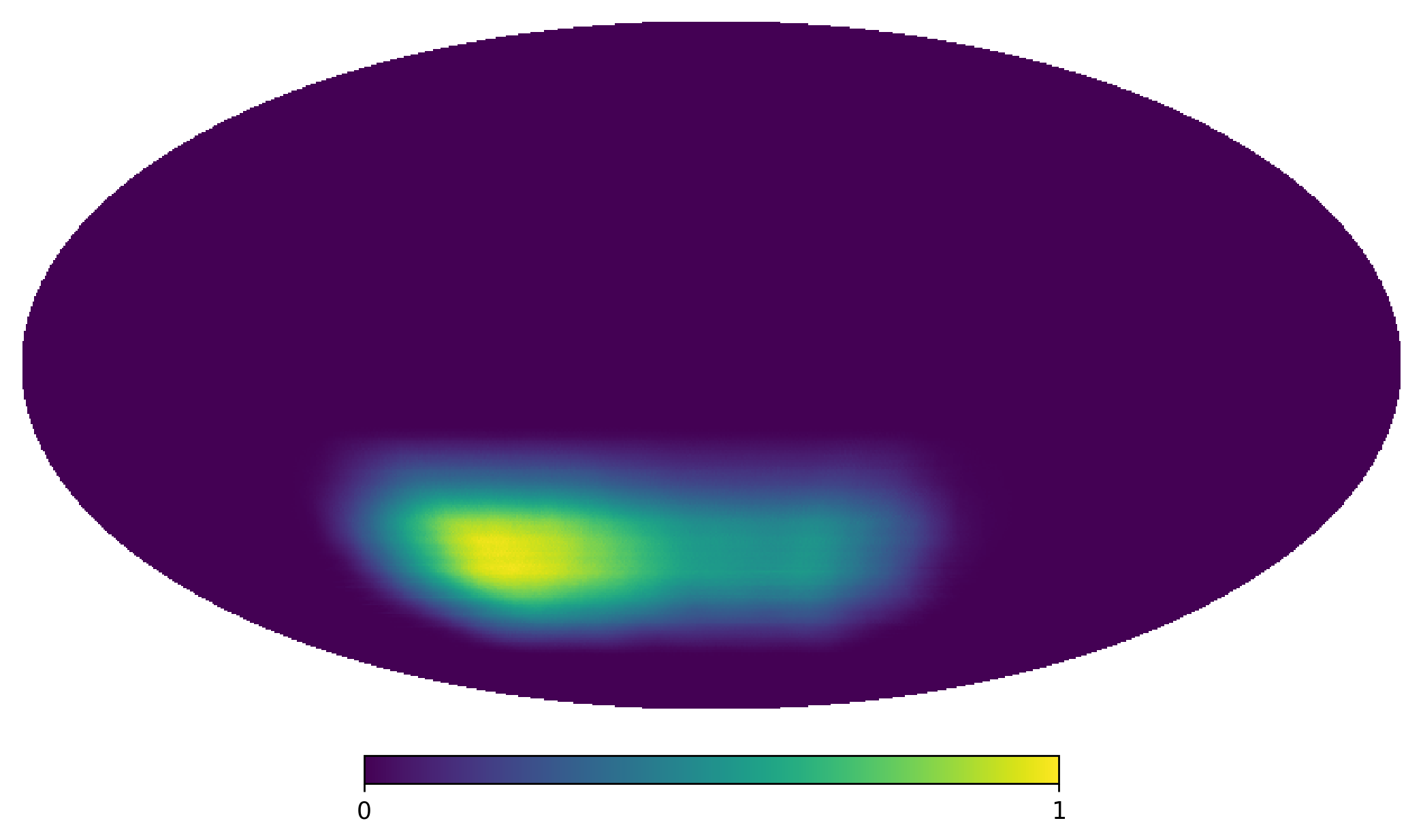}
    \caption{Normalised hits map, in Equatorial coordinates, for the full schedule we simulate in \textsc{toast}, which correspond to 10 observing days throughout a year totalling 139 hours. In this exercise, we only observe the South Galactic patch, while the real SO-SAT deep survey will also cover parts of the North Galactic patch.}
    \label{fig:hits_full}
\end{figure}

In total, our schedule has 167 observations for a total integration time of 139 hours. Each observation is almost an hour long and includes the 7 detector arrays, known as wafers, covering the focal plane of one SAT, for a total of $\sim 6000$ detectors per each f090/f150 band (however we use only one-eighth of these as described below). The normalised hits map of our full schedule is shown in Figure~\ref{fig:hits_full}, scanning the South Galactic patch. Our schedule is run in 8 sub-schedules that produce one filter+bin map each. Sub-schedules are used to balance between the flexibility to create splits of the data and computation time. These maps are linearly co-added.\footnote{Since the filtering and mapmaking operations are linear, coadding individual sub-schedule filter+bin maps yields the same map as applying the filter+bin mapmaker to the total schedule.} In particular, 5 sub-schedules correspond to observations when the South patch is rising through the sky (equivalent to 102 observations and 93.5 hours) and 3 sub-schedules when the South patch is setting in the sky (equivalent to 65 observations and 45.5 hours). This separation into sub-schedules is useful when defining splits for simulating a null test or trying to estimate cross-spectra without noise bias.

\subsection{Using \textsc{toast}} \label{sec:using_toast}

We simulate the observation of an input map by scanning it to time-ordered data (TOD) following a schedule like the one detailed above. Then observations are processed and mapped into filter+bin maps. In practice, this procedure is performed with the Time-Ordered Astrophysics Scalable Tools (\textsc{toast})\footnote{\url{https://github.com/hpc4cmb/toast} v3.0.0a25} software, as well as the \textsc{sotodlib}\footnote{\url{https://github.com/simonsobs/sotodlib} v0.5.0} library, which provides the necessary scripting and configuration to simulate specific detectors and telescopes for SO.

The TOD is simulated for every detector for a given telescope and detector wafer. The input map is scanned into a TOD, assigning, to each time sample, the value of the pixel that the telescope was pointing at (i.e. no interpolation is performed to avoid introducing an extra window function). Also, the TOD is modulated by the spinning of an idealised cryogenic continuously rotating HWP. The HWP is rotating near 2\,Hz, and therefore the polarised signal is modulated at 8\,Hz.

We then implement a series of analysis steps that would simulate the filtering in a CMB survey. To test the pipeline, we filter the simulated data more aggressively than we expect to do on the real SO data. The observations are processed in the following order:
\begin{itemize}
    \item A HWP synchronous signal filter is used to subtract the HWP synchronous signal introduced in the modulation of the TOD. This is a sines/cosines template fitted to data as a function of HWP angle and subtracted. The HWP filter is applied to one detector and one hour-long observation at a time.
    
    \item The TOD is demodulated by applying a bandpass filter centred at 8\,Hz and with width 4\,Hz and multiplying in Fourier space by the phasor $\exp(-4 i \chi(t))$, where $\chi(t)$ is the HWP angle as a function of time \citep{2007ApJ...665...42J,2024MNRAS.532.2309R}. The real part of the result is the demodulated $Q$ signal, while the imaginary part is the demodulated $U$ signal. Both are then low-pass filtered at 2\,Hz, since the bandpass filter will have already reduced all signals at higher frequencies. See~\cite{2007ApJ...665...42J,2014RScI...85c9901K,2016JLTP..184..534S,2017JCAP...05..008T,2024RScI...95b4504Y} for details.
    
    \item A ground filter is used to subtract the horizontal-coordinate-fixed contamination from the ground. This is a Legendre polynomial of order 10 fitted to data as a function of azimuth and subtracted. The ground filter is applied to one detector and one hour-long observation at a time.  No binning is needed: the polynomial templates are sampled at the boresight azimuth and projected out from the detector data.
    
    \item A high-pass filter is used to suppress the impact of low-frequency contamination such as the atmosphere and detector thermal drifts, among others. This is a Legendre polynomial of order 3 fitted to the data as a function of time and subtracted. This and the ground filter are applied simultaneously since they are templates deprojected in one matrix operation. The polynomial templates are projected out separately from each sub-scan (sweep across the field).
\end{itemize}
In Appendix~\ref{sec:individual_filters}, we illustrate the effect of individual filters. Finally, the TOD are binned into \textsc{healpix} $N_{\rm side}=512$ maps according to their pointing. We make some approximations to ease the computing cost of running these simulations for this work. We use a 40\,Hz sampling rate for scanning (real observations scan at 200\,Hz) and we also ``thin'' the focal plane by a factor of 8, i.e., only one out of every eight detectors is simulated. In Appendix~\ref{sec:fp_thinning} we discuss this in detail, showing how the impact is minimal while reducing the computation time by a factor of 8. The combined effect of the three data volume reductions (one day every month; 40\,Hz; 1/8th of detectors) is to reduce the computational requirements of the simulations by a factor of 1200 per frequency band per SAT. With our early data analysis, we have estimated that it takes $\sim 3$M core-hours to filter the TF simulations (300 in total, 100 realisations and pure-$T$, pure-$E$ and pure-$B$ modes for each) required for one year of data with one SAT and one band. This is avoiding any data thinning, which can lead to a significant speed-up without compromising the quality of the estimated transfer function. It is therefore entirely feasible to deploy the methods described here with the high-performance computing resources available to the collaboration.

\subsubsection{Time-domain noise simulations} \label{sec:noise}

While the section above describes an example of the filtering and processing steps applied to TOD, we also perform time-domain simulations of sources of noise and systematics. These sources are
\begin{itemize}
    \item \textbf{Atmospheric noise}: We simulate an unpolarised atmosphere~\cite{2011MNRAS.414.3272S,2020ApJ...889..120P} using an implementation of the model from Ref.~\cite{1995MNRAS.272..551C,2015ApJ...809...63E}. This is based on a 3D structure with Kolmogorov turbulence moving through the telescope beam, with a distribution of temperature, water vapour, and wind speed. This model was fitted to match {\sc Polarbear} observations at the Cerro Toco site. Within this model, and ignoring I-to-P leakage, this atmospheric emission does not impact the polarised signal, so the 1/f component of polarised noise is mostly controlled by the scanning speed and the choice of filters applied to the TOD. As such, this component has no effect on our $E,B$ fields. While the simulated unpolarised intensity maps are not used in this work, they will be useful for future analysis work.
    
    \item \textbf{Gaussian-distributed, uncorrelated white noise}: The detector white noise is based on the forecasted levels for an SAT given in the SO science paper~\cite{2019JCAP...02..056A}. In the \textsc{toast} instrument model, individual detectors are modelled such that a final 5-year deep survey map has noise levels of 2.6 and 3.3 $\mu$K-arcmin for f090 and f150, respectively. This corresponds to the baseline white map noise level in Ref.~\cite{2019JCAP...02..056A}.
    
    \item \textbf{Scan Synchronous Signal (SSS)}:
    We use an ad hoc model where the telescope scans a fixed map in horizontal coordinates to emulate both optical pickup and signals internal to the instrument (e.g. electromagnetic interference from the azimuthal drive).  All detectors observe the same map, but the map is drawn separately for each simulated observation.  The map itself is Gaussian white noise, convolved with a $1^\circ$ beam, with amplitude inversely proportional to the observing elevation and scaled to 1\,mK RMS at $45^\circ$ observing elevation. The elevation dependence is added to reflect the fact that far-side lobe pickup intensifies at lower observing elevations. The $1^\circ$ smoothing is chosen to avoid sharp features in the simulated signal. Finally, the overall amplitude of the SSS is chosen to overwhelm CMB signal for the sky modes that are degenerate with it.  These choices are anecdotally supported by {\sc polarbear} data but are not based on rigorous analysis.  The model will be thoroughly calibrated against real data in the future.
    
    \item \textbf{Half-Wave Plate Synchronous Signal~(HWPSS)}: This is added to the TOD from a template constructed in Ref.~\cite{2018SPIE10708E..48}.
\end{itemize}
The noise simulations are run using the same schedule as described in Section~\ref{sec:schedule}, and noise levels are scaled to match the SO-SAT expected noise level after a 1-year deep survey (2 telescopes and~6000 detectors on each for 4170 hours)~\cite{2019JCAP...02..056A}.

Our overall noise model lies between the optimistic ($\ell_{\rm knee}=25$) and pessimistic ($\ell_{\rm knee}=50$) SO-SAT performance levels~\cite{2019JCAP...02..056A}. To evaluate this, we take the pure noise simulations, deconvolve the filtering with the corresponding transfer function, and fit a noise curve model
\begin{equation}
    N_{\ell} = N_{\rm white}\left[ 1+(\ell / \ell_{\rm knee})^{\alpha_{\rm knee}} \right]    
\end{equation}
to the mean of 10 only-noise f090 realisations, finding $\ell_{\rm knee}=44 \pm 4$ and $\alpha_{\rm knee}=-3.22 \pm 0.15$. We generate 100 realisations of noise observations for the f090 frequency channel and the same for the f150 channel.

\subsubsection{Observation matrix} \label{sec:obsmat}
As described in Section~\ref{sec:per_ell}, the per-$\ell$ estimation method requires thousands of \textsc{toast} simulations. The observation matrix $\matr{R}$ is such that a biased (and masked) observed map $\tilde{\vect{m}}$ is expressed as in Equation~\eqref{eq:obs_matrix}. $\matr{R}$ has a shape $3N_{\rm pixels}$ by $3N_{\rm pixels}$ (to capture the three Stokes components $I$, $Q$, and $U$). $\matr{R}$ can be calculated explicitly \cite{2016ApJ...825...66B} with 
\begin{equation}
    \matr{R} = (\matr{P}^T \matr{N}^{-1} \matr{P} )^{-1} \matr{P}^T \matr{N}^{-1} \matr{Z} \matr{P} \text{,}
\end{equation}
where we recognise the elements familiar from mapmaking, discussed in Section~\ref{sec:filterbin}, and $\matr{Z}$ is the time-domain filtering operator, which includes all the operations done to an observation. It is a square matrix that has the dimensions of the time-ordered data and performs linear regression of time domain templates represented as columns of matrix $\matr{F}$:
\begin{equation}
    \matr{Z} = \mathbf{1} - \matr{F}\left(\matr{F}^T\matr{N}^{-1}_Z\matr{F}\right)^{-1}\matr{F}^T\matr{N}^{-1}_Z.
\end{equation}
In place of a noise covariance matrix, $\matr{N}$, the filtering operator uses a binary diagonal matrix, $\matr{N}_Z$, that can mask out data we do not want affecting the template amplitudes: typically samples drawn near planets, point sources or intense Galactic emission.

The filtering matrix, $\matr{Z}$, operating on a TOD vector, $\vect{d}$, can be understood to perform three separate operations:
\begin{enumerate}
    \item Perform a generalized least squares fit of $\vect{d}$, in terms of the time domain templates that form the columns of $\matr F$.  This results in a vector of template amplitudes: $\vect{a} = \left(\matr{F}^T\matr{N}^{-1}_Z\matr{F}\right)^{-1}\matr{F}^T\matr{N}^{-1}_Z \vect{d}$.
    \item Unroll (weight and co-add) the templates, $\matr F$, according to the fitted amplitudes, $\vect{a}$, to produce a model of the TOD: $\vect{d}_\mathrm{model} = \matr F \vect{a}$.
    \item Subtract the TOD model from the input to produce a filtered version of the input TOD: $\vect{d}_\mathrm{filtered} = \vect{d} - \vect{d}_\mathrm{model}$.
\end{enumerate}
\section{Results} \label{sec:results}

In this section, we show the measured band power window functions (BPWFs) using the two methods presented in Section~\ref{sec:maps_to_spectra} applied to the filtered maps as observed with the simulated schedule described in Section~\ref{sec:schedule}. These can be used to deconvolve the effect of mode-mixing and reconstruct the power spectra of the CMB. We show the transfer function (TF) method adequately recovers the input power spectrum over the full multipole range of interest. Finally, we discuss limitations of the TF method.

\subsection{Comparison of band power window functions} \label{sec:bpwf}

For both the per-$\ell$ estimation method and the TF method, we filter simulations with the observation matrix, which in practice is multiplying $\matr{R}$ with the input map. Since we need many simulations to constrain the BPWFs, especially those associated with polarisation leakage, we simplify our analysis and calculate band power bins up to $\ell_{\rm max}=255$ with width $\Delta \ell = 10$ for both methods. For the numerical per-$\ell$ method, we generate and filter full-sky simulations that excite every individual $\ell$ between 2 and $\ell_{\rm max}$. The estimation for each multipole is run with 6144 realisations.

For the TF method, we filter our power-law simulations with the observation matrix, so we can guarantee the exact same filtering operations between both methods. We run the estimation of the TF with 100 realisations of a power-law with spectral index $-2$ and the same binning scheme.

\begin{sidewaysfigure}
    \centering
    \includegraphics[width=1\textwidth]{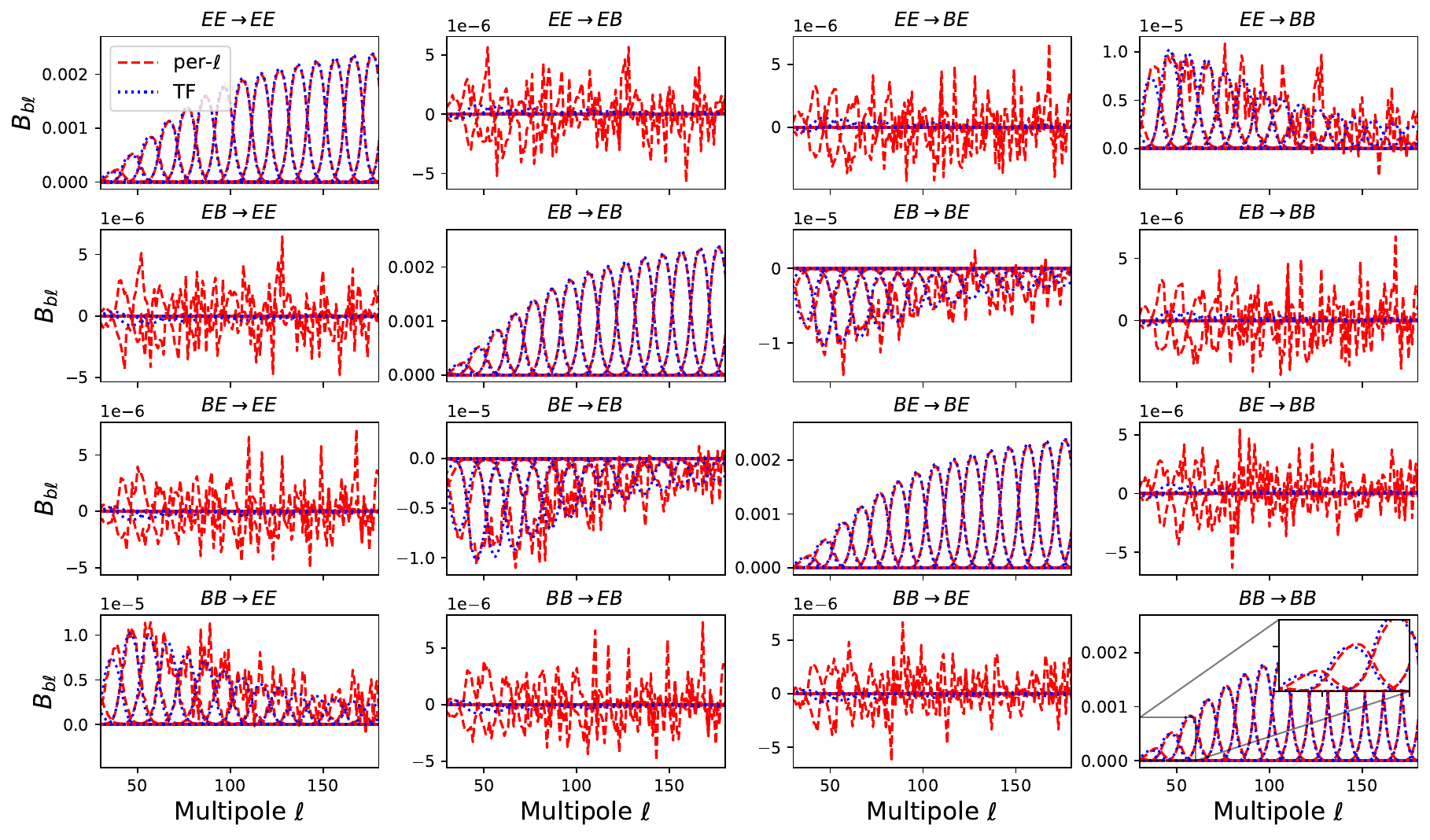}
    \caption{Reconstructed BPWFs for the full schedule and for the binning scheme $\ell=30-255$ with size $\Delta \ell = 10$ (in the plot we limit to $\ell=30-180$ for clarity). The titles above each subplot refer to the indices $\alpha'\beta' \to \alpha\beta$ in Equations~\eqref{eq:cl_bbl}, \eqref{eq:bbl_mask}, and \eqref{eq:bbl_obs}. Red-dashed is the numerical per-$\ell$ solution estimated with 6144 simulations per $\ell$. Blue-dotted is the TF approximation estimated from 100 power-law simulations. In the $BB \to BB$ panel, we show a zoom-in inset highlighting the difference between the two methods at low-$\ell$. This is for the f090 frequency channel.}
    \label{fig:bpwf_comparison}
\end{sidewaysfigure}

The BPWFs $B_{\alpha \beta b}^{\alpha' \beta' \ell}$, shown as $4 \times 4 = 16$ $B_{b \ell}$ combinations of $\alpha' \beta' \to \alpha \beta$ polarisations and for band power $b$ and multipole $\ell$ are shown in Figure~\ref{fig:bpwf_comparison}. Most band powers $b$ are plotted together (we limit to $\ell=30-180$ for visualisation). The numerical per-$\ell$ results are shown as dashed red lines for the band powers, while the TF results are shown as dotted blue for the same binning scheme.

First, we note the power suppression arising from observing and filtering in our mapmaking. The auto-BPWFs (the diagonal panels in Figure~\ref{fig:bpwf_comparison}) show excellent agreement between both methods in the higher-$\ell$ band powers. This implies that the approximation of Equation~\eqref{eq:mcm_tf} is valid, for example, at the level of $\sim 1$\% at $\ell \sim 200$. In other words, the additional mixing between different angular scales caused by filtering is subdominant compared to that caused by the mask, given the steepness of the power spectra under study. However, at lower multipoles ($\ell<100$), where the suppression is larger, there is a clear difference, highlighted by the zoom-in inset in the lower right panel. The TF method assumes that the suppression of modes is the same at all multipoles within a given band power. This assumption breaks down at larger angular scales, resulting in the difference observed in Figure~\ref{fig:bpwf_comparison} at lower multipoles.

The relevant cross-BPWFs (e.g., the $EE \to BB$ panel in Figure~\ref{fig:bpwf_comparison}, which shows the $E$-mode leakage into $B$-modes, or $EE \to EB$, which is even smaller and noisier) are ${\sim} 50-250$ times smaller than the auto-BPWFs. Due to the signal-to-noise ratio, where the noise is dominated by the cosmic variance of a map realisation, this is difficult to constrain with fewer simulations for the numerical per-$\ell$ method. Hence, we need thousands of simulations and the per-$\ell$ method is limited by how much we are willing to shrink the cosmic variance. The reconstruction is noisy as can be seen in the figure. This means that the reconstruction of $B$-modes with the numerical per-$\ell$ method might be noisy or biased due to contamination from $E$-modes, since in general for the CMB the $B$-modes are a few orders of magnitude smaller than the $E$-modes, and the leakage might represent a significant fraction of the signal.

\begin{figure}
    \centering
    \includegraphics[width=1\textwidth]{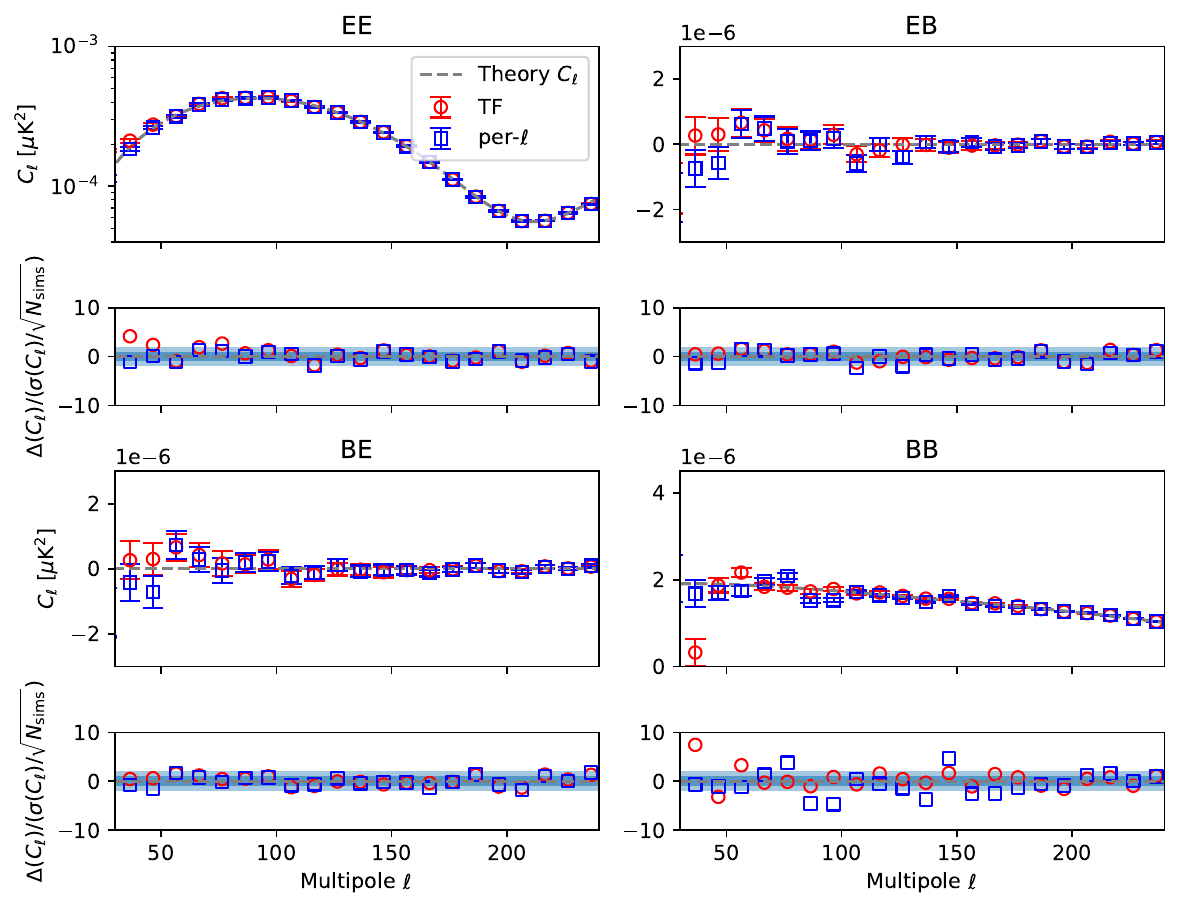}
    \caption{
    Comparison of the reconstructed polarisation spectra for the f090 channel using CMB simulations with only lensing $B$-modes. The four panels are in clockwise order the $EE$, $EB$, $BB$, $BE$ spectra. The spectra are the mean over $N_{\rm sims}=100$ realisations, while the error bar is the standard deviation divided by $\sqrt{N_{\rm sims}}$. The theory CMB spectra are shown as the dashed grey lines and have been convolved with the corresponding Gaussian beam. The smaller panel at the bottom of each spectrum shows the residual between the average of the reconstructed and theory spectra, $\Delta (C_{\ell})$, divided by $\sigma(C_{\ell})/\sqrt{N_{\rm sims}}$. The shaded areas correspond to $\pm 1$ and $\pm 2 \sigma$. The recovered spectra are visibly biased at $\ell<70$ in this noiseless case. }
    \label{fig:reconstructed_cmb_comparison}
\end{figure}

We also simulate pure CMB realisations, generated with the \textit{Planck} best-fit cosmology~\cite{2020A&A...641A...6P}, including $B$-modes arising only from gravitational lensing ($r=0$), as described in Section~\ref{sec:input_simulations}. We generate them at $N_{\rm side}=128$ and filter them with the same scheme represented by the observation matrix. The reconstructed polarisation spectra with both methods are shown in Figure~\ref{fig:reconstructed_cmb_comparison}. The four square panels show the $EE$, $EB$, $BB$, and $BE$ spectra in clockwise order. In each, we show the mean $C_{\ell}$ across $N_{\rm sims}=100$ realisations. The error bar is the error of the mean, i.e. the standard deviation across realisations divided by $\sqrt{N_{\rm sims}}$, and as such, these $\sigma$ values are 10 times smaller than $\sigma$ for a single realisation. The thinner lower panel on each spectrum shows the ratio between spectrum residuals and the associated error. The residual $\Delta(C_{\ell})$ is the difference between the measured decoupled power spectrum and the fiducial theory spectrum binned using the BPWFs. The observed CMB maps include beam smoothing, and so the theory spectra are convolved with a $30^\prime$ Gaussian beam. Since these are pure CMB realizations, the only source of variance is cosmic variance.

The $EE$ spectrum is affected the least by contamination due to leakage, since $B \to E$ leakage is very small. For example, the first bin in the $EE$ deconvolved spectrum (upper left corner, $\ell = 30-40$) shows a ${\sim}4 \sigma$ bias for the TF method, while per-$\ell$ shows mostly unbiased band powers. This arises because the power spectrum of the TF simulations with power-law index $-2$ varies largely within the first multipole bin and does not match well the shape of a noiseless CMB power spectrum, leading to a bias in the TF estimate. However, in Section~\ref{sec:null_tests} we will show that this bias is not important in the context of studying the precision of the 1-year SO SAT deep survey discussed in this paper. Another issue to note is that the cross-BPWFs for the per-$\ell$ method are noisy, so the leakage estimation is noisy, and if leakage is relevant, then e.g., the $BB$ deconvolved spectrum will be vulnerable to large contamination from $E \to B$ leakage. This is very noticeable in the $BB$ spectrum on the lower right panel of Figure~\ref{fig:reconstructed_cmb_comparison}, which shows large dispersion by several sigma due to $E \to B$ leakage. Finally, the $EB$ and $BE$ spectra are the same in the case of the TF method, since the TF is calculated from the same realisation, but different for the per-$\ell$ method, where different realisations for $EB$ and $BE$ are used. In both cases, the residuals are consistent with zero.

The TF method can reconstruct all deconvolved power spectrum bins between $\ell=30$ and 250 from biased filter+bin maps through $2 \times \mathcal{O}(100)$ simulations with a bias below 1$\sigma$ (for a single cosmic-variance-limited CMB simulation).\footnote{We note that Figure~\ref{fig:reconstructed_cmb_comparison} shows the bias on the mean of 100 CMB-only sky realisations, which must be divided by $\sqrt{N_{\rm sims}}=10$ to yield the bias on a single realisation.} This is acceptable, especially given that real observations will contain additional noise. By comparison, the numerical per-$\ell$ method is impractical to run since it requires thousands of simulations per multipole (which is virtually impossible), or constructing an observation matrix every time that the filtering and mapmaking procedure is updated. In Section~\ref{sec:null_tests}, we will illustrate how the biases induced by the approximations made by the TF method are sub-dominant to the noise, and as such not important, through the use of null tests.
Hence, we will adopt the TF for the rest of our analysis.

\subsection{Full multipole range using the transfer function method} \label{sec:tf_full_multipole}

\begin{figure}
    \centering
    \includegraphics[width=1\textwidth]{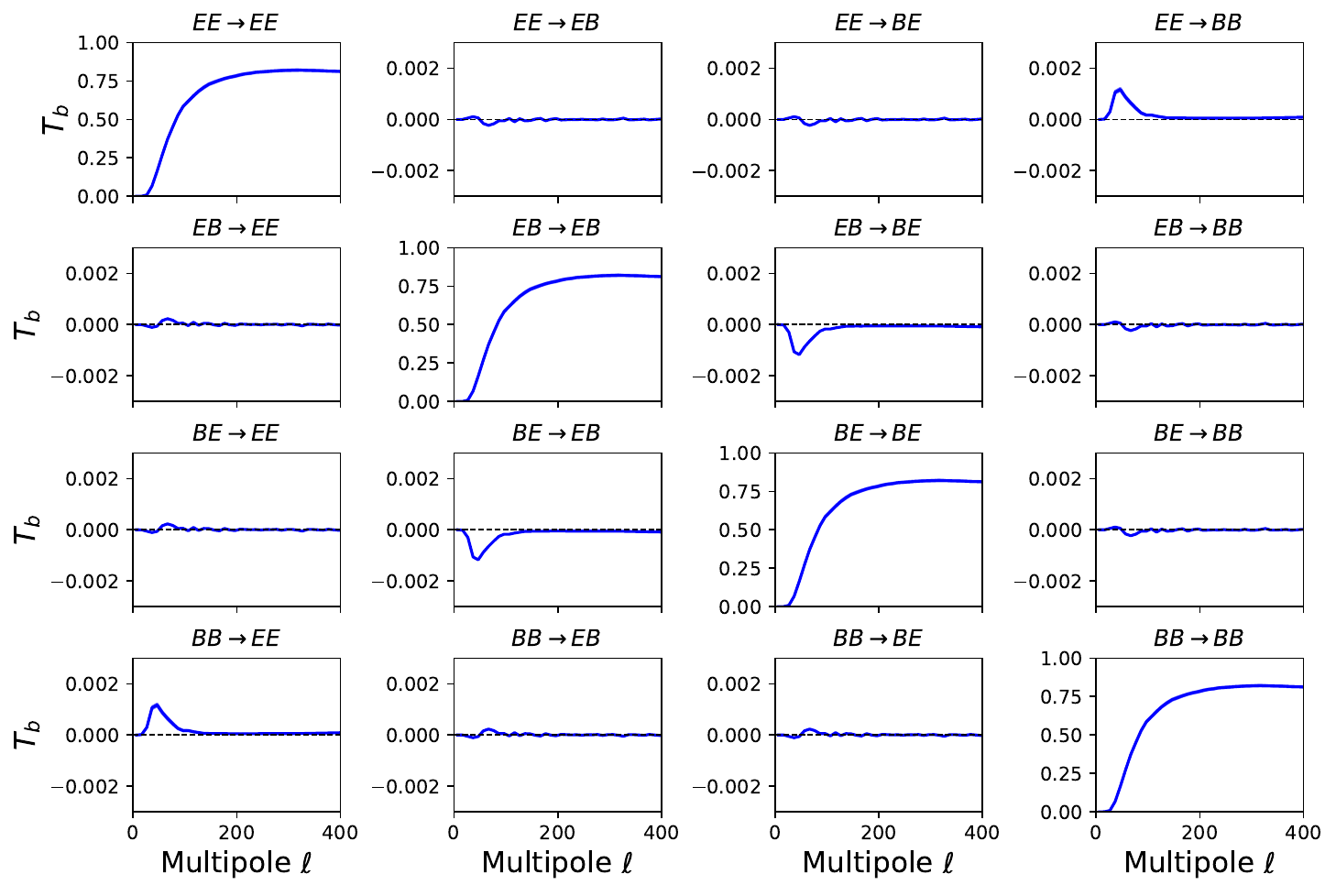}
    \caption{
    The transfer function for the f090 channel, showing the 16 $\alpha' \beta' \to \alpha \beta$ polarisations estimated with Equation~\eqref{eq:tf_least_squares}, averaged over 100 realisations of power-law simulations. The shaded area (barely thicker than the blue line) represents the $\pm 1$ standard deviation divided by $\sqrt{N_{\rm sims}}=10$. This transfer function is estimated using more aggressive filtering than we expect to use in the real SO data analysis.
    }
    \label{fig:tf_cmb}
\end{figure}

In order to evaluate the TF method across the full SAT multipole range ($\ell = 30-300$), we run timestream-to-map filtering on $N_{\rm side}=512$ simulated input maps. We remind the reader that estimating the TF only requires a modest number of simulations, which can therefore be filtered using brute-force methods. This is in contrast with the per-$\ell$ method, where the number of simulations is significantly larger, and memory-intensive observation matrices must be used to enable fast observation and filtering. Figure~\ref{fig:tf_cmb} shows the transfer function $T_b^{(\alpha \beta)(\alpha' \beta')}$ for the full observation schedule described in Section~\ref{sec:schedule} calculated from 100 realisations. Again, we note that this transfer function is estimated using more aggressive filtering than we expect to use in the real SO data analysis. The auto-terms $\alpha\alpha \to \alpha\alpha$ in the diagonal of the figure show the suppression at larger scales due to filtering, and then plateau at ${\sim}0.8$. In Appendix~\ref{sec:individual_filters}, we illustrate why this happens, where the high-pass polynomial filter of order 3 is most responsible.

\begin{figure}
    \centering
    \includegraphics[width=1\textwidth]{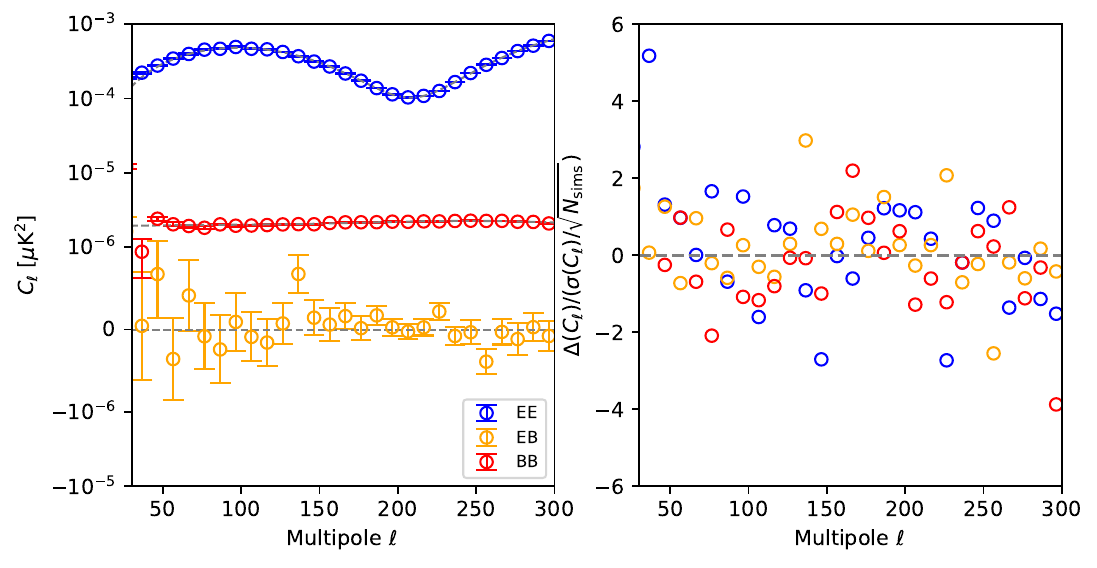}
    \caption{
    Deconvolved spectra for the same CMB shown in Figure~\ref{fig:reconstructed_cmb_comparison} for the full schedule and using the TF method. The fiducial $EE$ and $BB$ spectra are shown in grey dashed lines. We can achieve unbiased power spectra estimates for the range $\ell=40-300$ at moderate computational cost. This is for the f090 frequency channel.
    }
    \label{fig:Cell_cmb}
\end{figure}

The TF-deconvolved spectra for the observed CMB and for the full range $\ell=30-300$, are shown in Figure~\ref{fig:Cell_cmb}. For each spectrum we show the mean $C_\ell$ with error on the mean across $N_{\rm sims} = 100$ realisations. The right-hand side panel shows the residuals divided by the error bar. No systematic bias in the residual is visible, except for the first band power $\ell=30-40$. In Section~\ref{sec:null_tests}, we will show that this bias is sub-dominant and therefore not important when noise is added to these pure CMB simulations, through the use of null tests.

\subsection{Limitations of the transfer function method} \label{sec:tf_limitations}

To illustrate the limitations of the TF method, we use the same TF estimated in Section~\ref{sec:tf_full_multipole} to reconstruct a different fiducial spectrum with a very different shape, a steep spectral index $\alpha_{\rm s}=-6$, which has large variation across each of the band powers. This case is not representative of what the true sky is expected to look like, but this exercise is to push the boundaries of the method.

\begin{figure}
    \centering
    \includegraphics[width=1\textwidth]{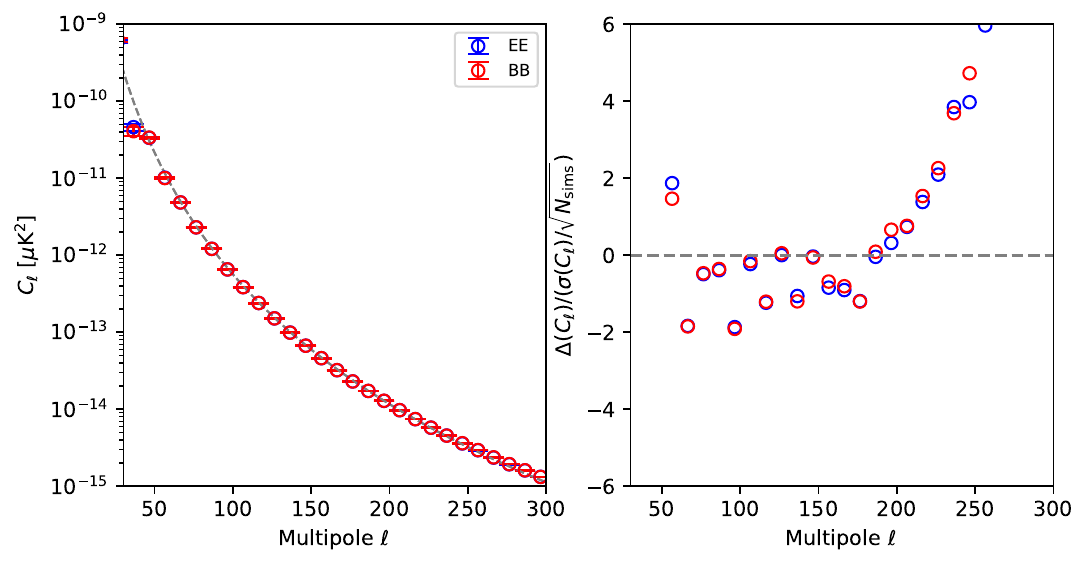}
    \caption{
    Deconvolved spectra for a power-law theory spectrum with a steep index $\alpha_{\rm s}=-6$. This uses 100 realisations, and the full schedule. We use the same TF shown in Figure~\ref{fig:tf_cmb}. This figure follows the same convention as Figure~\ref{fig:Cell_cmb}. This is for the f090 frequency channel.
    }
    \label{fig:Cell_PL_minus6}
\end{figure}

We filter 100 power-law realisations with the same fiducial $C_{\ell}$ in the $EE$ and $BB$ spectra, with $\alpha_{\rm s}=-6$ spectral index and the same $30^\prime$ beam smoothing. Figure~\ref{fig:Cell_PL_minus6} shows the reconstructed spectra and the bias in the residual, analogous to Figure~\ref{fig:Cell_cmb}. The entire multipole range in the spectra seems to be biased. The coupling of modes induced by observing/filtering within a band power bin is large enough that the TF is no longer a good approximation. 
\section{Null test} \label{sec:null_tests}

In this section, we assess the relative bias of power spectra reconstructed from simulated data subsets (``splits'') that have undergone significantly different filtering, and hence possess different transfer functions. If the power spectrum reconstruction is unbiased and robust against different filtering prescriptions, the power spectrum corresponding to the difference map should be statistically consistent with noise. First, we take a look at pure signal with only CMB, and then we add realistic noise.

\subsection{Definition of splits}

\begin{figure}
    \centering
    \includegraphics[width=0.80\textwidth]{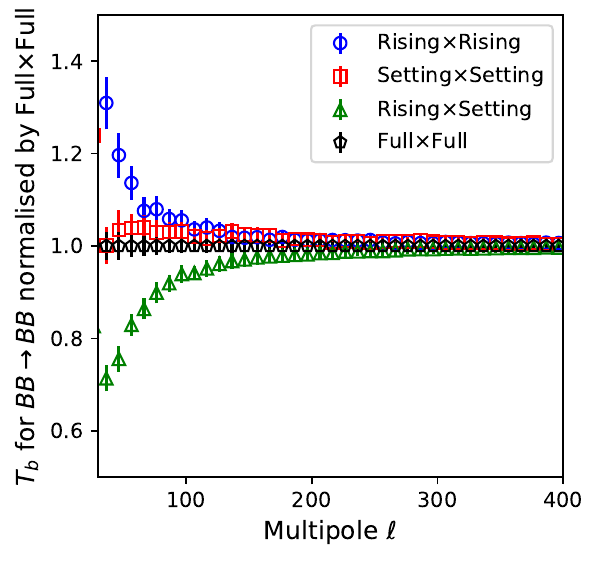}
    \caption{
    The TF for $BB \rightarrow BB$ for the case of using rising and setting scan splits. All are normalised by the Full$\times$Full TF. We also include the full schedule in black pentagons, the same as shown in Figure~\ref{fig:tf_cmb} but in a more restricted mask. We note how all combinations are noticeably different from each other. This is estimated from 100 power-law realisations. The error bars are the standard deviation across the realisations divided by $\sqrt{100}$. This is for the f090 frequency channel.
    }
    \label{fig:tf_cmb_splits}
\end{figure}

In order to maximize the potential difference between transfer functions, we minimize cross-linking\footnote{Cross-linking refers to the technique of scanning sky coordinates from multiple directions to mitigate potential systematics.} in the data splits, partitioning the scans into rising (R) and setting (S), and calculating the different TFs for each map combination: R$\times$R, S$\times$S, R$\times$S, and S$\times$R where $\times$ represents cross-correlation between two maps. Even further, those TFs will be different from using the full mission, labelled Full and corresponding to R$+$S. This is illustrated in Figure~\ref{fig:tf_cmb_splits}, where we show the $BB \rightarrow BB$ TF for R$\times$R, S$\times$S, R$\times$S, and Full$\times$Full, normalised by Full$\times$Full.

\begin{figure}
    \centering
    \includegraphics[width=0.325\textwidth]{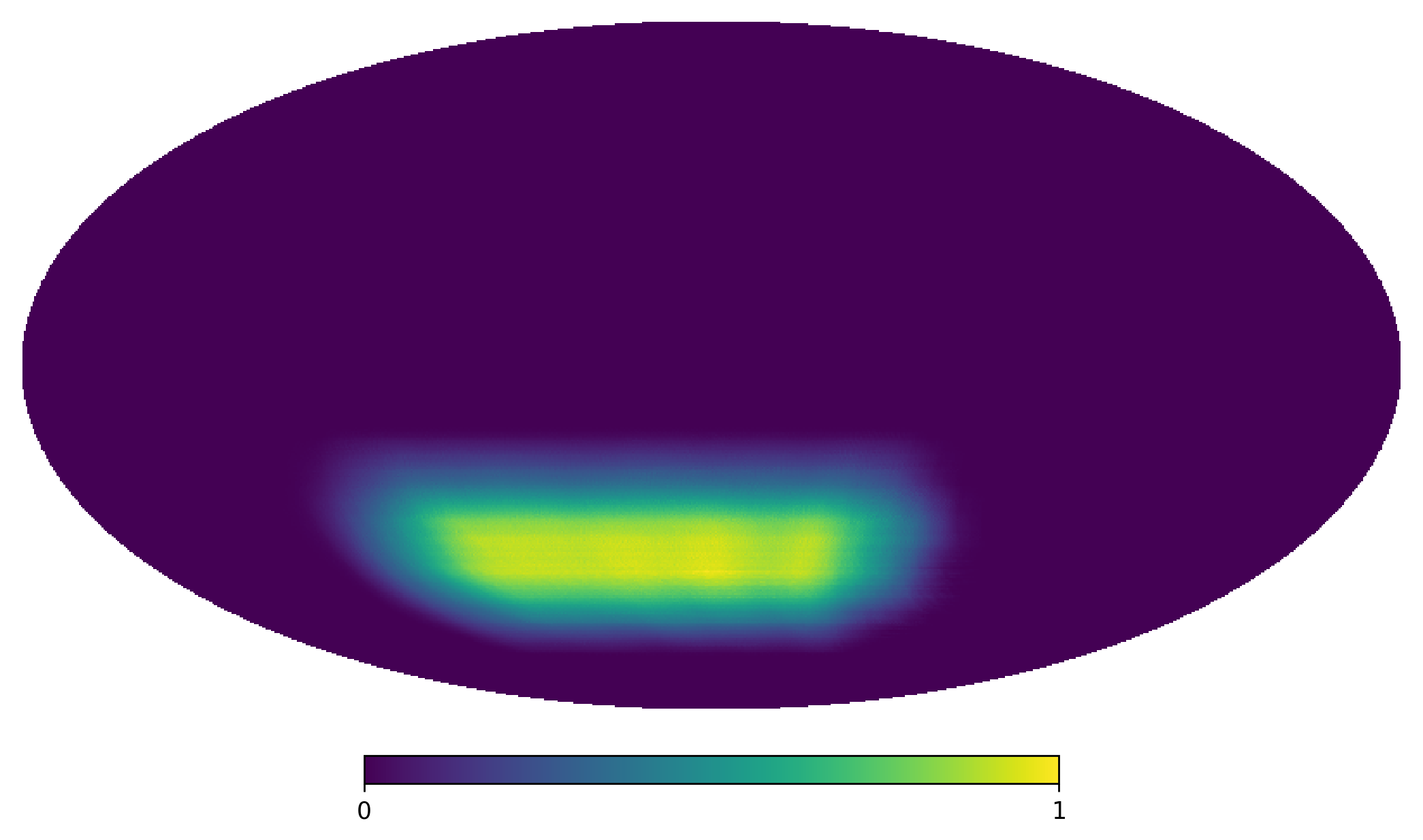}
    \includegraphics[width=0.325\textwidth]{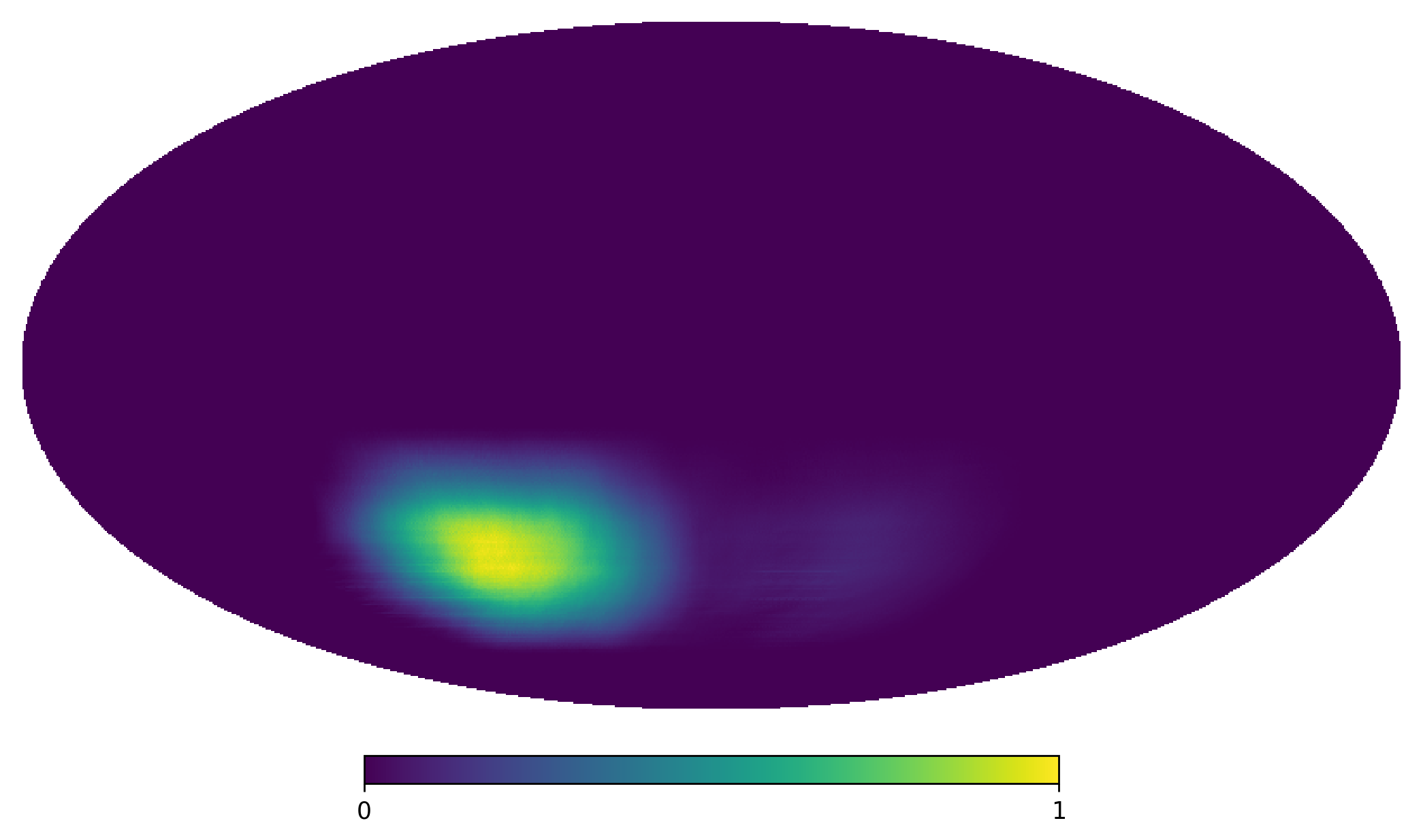}
    \includegraphics[width=0.325\textwidth]{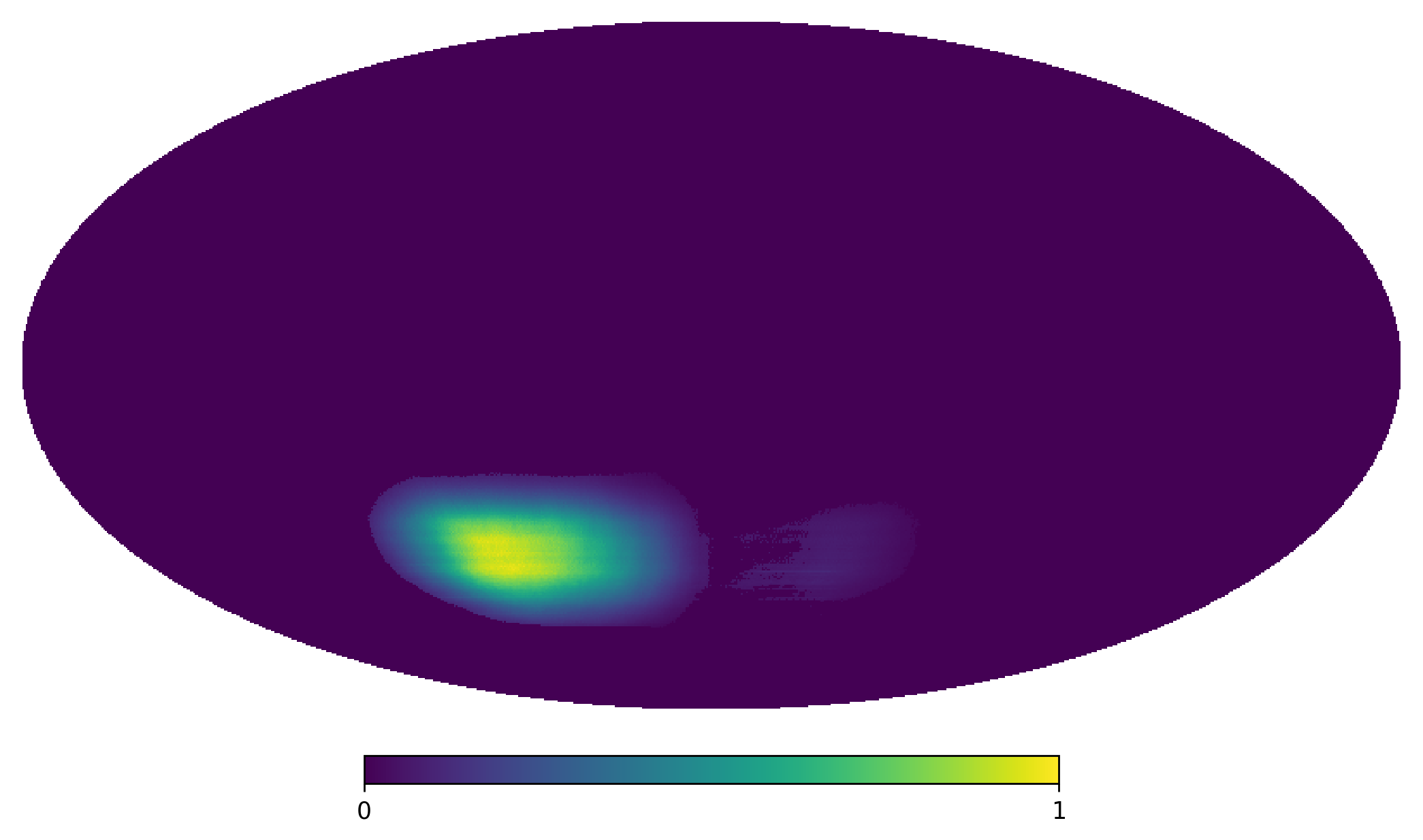}
    \caption{Normalised hits maps for splits. Left, normalised hits map for the rising scans. Centre, normalised hits map for the setting scans. Right, the common hits map is constructed from the intersection of the rising and setting hits maps. We build a mask from this hits map to estimate power spectra in Section~\ref{sec:null_tests}. All maps are shown in Equatorial coordinates.
    }
    \label{fig:masks_splits}
\end{figure}

In Figure~\ref{fig:masks_splits}, we show the normalised hits maps for the rising and setting scans at the left and centre, respectively. As a result of the scanning strategy used here, the setting scans observe approximately one-half of the South patch, while the rising scans almost uniformly cover that region. We construct a common hits map using the intersection of both obtained by multiplying the rising and setting hits maps, but keeping only the pixels where in both cases the hits count is greater than the median computed over all pixels. This common hits map is shown on the right-hand side of the figure. Finally, we build a mask from this hits map, which we use for power spectra calculation in this section.

\subsection{CMB-only test} \label{sec:null_test_pure_signal}

First, we calculate the residual spectra for 100 CMB simulations with only lensing $B$-modes. Since we only have signal in our maps, we can calculate all spectra via direct correlation of the maps. We calculate the R$\times$R, S$\times$S, and R$\times$S spectra (S$\times$R in this case is the same as R$\times$S). For each type of spectra, we use the corresponding TF calculated from co-adding the pure power-law simulations in the same split combination. The residual spectrum $\Delta C_{b}^{XY}$ for the null test is given by 
\begin{equation}
\Delta C_{b}^{XY} = C_{b}^{X_{\rm R} Y_{\rm R}} + C_{b}^{X_{\rm S} Y_{\rm S}} - C_{b}^{X_{\rm R} Y_{\rm S}} - C_{b}^{X_{\rm S} Y_{\rm R}} \text{,}
\end{equation}
where $X,Y \in E,B$ and $b$ indexes band power bins. We calculate this residual for each of the 100 CMB realisations. Then, we define
\begin{equation} \label{eq:chi2}
  \chi^2(XY) = \sum_{b} \left[ \Delta C_b^{XY} / \sigma(\Delta C_b^{XY}) \right] ^2 \text{,} 
\end{equation}
where $\sigma(\Delta C_b^{XY})$ is the standard deviation across the 100 realisations. In this case of pure signal, the statistical uncertainties are dominated by residual cosmic variance caused by differential filtering. We note that this equation ignores uncertainties correlated across different angular scales, effectively assuming a diagonal covariance matrix. We adopt this approximation because the covariance matrix estimated from only 100 simulations is too noisy to invert reliably. The matrix itself is very close to diagonal, with all off-diagonal elements consistent with zero. The sum over $b$ goes over 27 bins in the range $\ell = 30-300$ with width $\Delta \ell = 10$.

\begin{figure}
    \centering
    \includegraphics[width=0.49\textwidth]{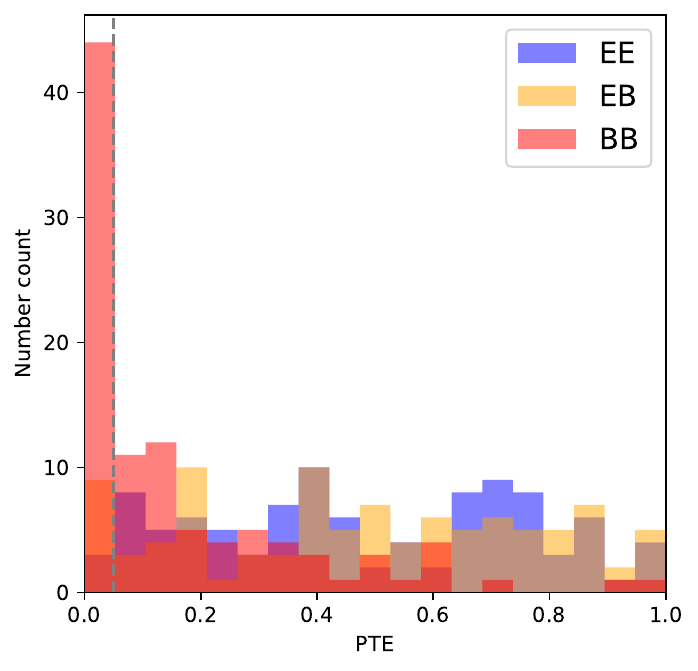}
    \includegraphics[width=0.49\textwidth]{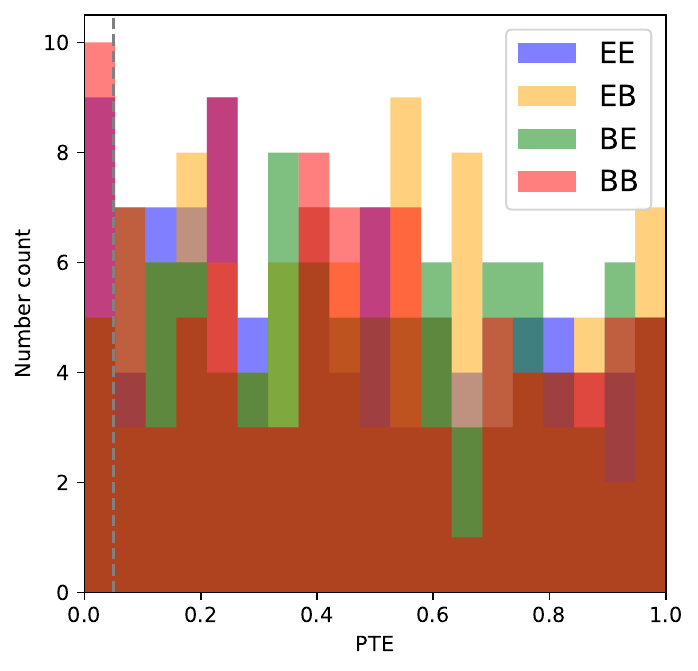}
    \caption{
    Histograms of PTEs for the null tests. This is for the residual spectra R$\times$R+S$\times$S-R$\times$S-S$\times$R, where R and S represent rising and setting scans splits. (Left) Null test for the pure CMB signal. The $BB$ spectrum in this case is very biased and a clear fail of the null test. (Right) Null test for a realistic survey, which includes noise and foregrounds on top of CMB. Each $R$ or $S$ split is further split into 2 half missions and the spectra are the cross between half missions to avoid a noise bias. This is for the f090 frequency channel. The dashed grey line marks PTE=0.05.
    }
    \label{fig:null_tests}
\end{figure}

If the residuals are consistent with noise, the $\chi^2$ statistic of Equation~\eqref{eq:chi2} should be consistent with a chi-squared distribution with 27 degrees of freedom (DOF), this being the number of band powers in the range $\ell=30-300$. We can calculate the probability-to-exceed (PTE) of each realisation. The histogram of the 100 PTEs is shown on the left-hand side of Figure~\ref{fig:null_tests} for the three $XY$ polarisations: $EE$, $EB$, and $BB$. If the residual in this null test is consistent with noise, we would expect a uniform distribution of PTEs between 0 and 1.

\begin{table}
    \centering
    \caption{PTEs for the uniform distribution K-S test over a sample of 100 realisations.
    \label{table:null_tests}
    }
    \begin{tabular}{l l l l l}
        \hline
        \hline
        Null test & $EE$ & $EB$ & $BE$ & $BB$ \\
        \hline
        pure signal & 0.514 & 0.855 & -- & $3 \times 10^{-27}$ \\
        realistic with noise & 0.172 & 0.421 & 0.912 & 0.152 \\
        \hline
    \end{tabular}
\end{table}

We perform a two-sided Kolmogorov-Smirnov (K-S) test, comparing our sample of PTEs to a uniform distribution.\footnote{We use the implementation from \textsc{scipy} \url{https://docs.scipy.org/doc/scipy/reference/generated/scipy.stats.kstest.html}} The PTE of each K-S test is listed in the first row of Table~\ref{table:null_tests}. We pass the null test in the case of the $EE$ and $EB$ spectra, but we can reject to a very high significance the null hypothesis for the $BB$ spectrum, which is clearly not a uniform distribution as shown in the left side panel of Figure~\ref{fig:null_tests}. As described in Section~\ref{sec:bpwf}, the TF is an approximation and the BPWF reconstruction can be biased. This null test failure will disappear when realistic noise is added to the simulation, as shown in the second row of Table~\ref{table:null_tests}, and explained below.

\subsection{Signal and noise test} \label{sec:null_test_signal_noise}

Our survey will not be pure CMB signal but also contain anisotropic and non-Gaussian polarised foregrounds, as well as noise. We include the \textsc{pysm} \texttt{d1} and \texttt{s1} foregrounds. The main issue of concern for our analysis is the spatial anisotropic and non-Gaussian nature of these templates, which could potentially induce biases in the deconvolved spectra due to a TF calculation that relies on spatially isotropic Gaussian simulations, as described in Section~\ref{sec:input_simulations}. We use these models with moderate levels of SED variability as our main interest is to assess the impact on mode coupling from the highly spatially anisotropic and non-Gaussian foregrounds. Using models with more SED complexity or even frequency decorrelation is beyond the scope of our paper. We leave the detailed study of foreground-induced biases in a TF reconstruction for future work.

For noise, we use the realisations scaled to a 1-year deep survey with noise $\sim 5$\,$\mu$K-arcmin for f090. We can co-add CMB, foregrounds, and noise in the same rising and setting scan splits, but we would be subjected to noise bias. To obtain noise-unbiased estimates of the signal power spectra, we further split each rising or setting map into a half-mission (HM) 1 and 2 split and cross-correlate between them. Since HM splits do not have significantly different hits maps, we continue the use of the mask from Section~\ref{sec:null_test_pure_signal}. In this case, the residual spectrum is given by
\begin{equation}
\Delta C_{b}^{XY} = C_{b}^{X_{\rm R, HM1} Y_{\rm R, HM2}} + C_{b}^{X_{\rm S, HM1} Y_{\rm S, HM2}} - C_{b}^{X_{\rm R, HM1} Y_{\rm S, HM2}} - C_{b}^{X_{\rm S, HM1} Y_{\rm R, HM2}} \text{.}
\end{equation}
Note that in this case the $EB$ and $BE$ null tests are different. In any realistic scenario such as this, the residual from a null test should be dominated by statistical noise rather than cosmic variance.

We follow the same procedure as in Section~\ref{sec:null_test_pure_signal}. The histogram of PTEs for the chi-squared distribution for 100 realisations is shown on the right-hand side of Figure~\ref{fig:null_tests}. We perform a K-S test on each of the 4 $EE$, $EB$, $BE$, and $BB$ polarisations, whose PTE is shown in the second row of Table~\ref{table:null_tests}. In this case, we pass the null test for $BB$ and we can state that our rising and setting splits are statistically compatible. This means that the TF method can be used for the reconstruction of noise-unbiased power spectra for filtered maps, at a $\sim 5$\,$\mu$K-arcmin noise level.
\section{Full time-domain simulations} \label{sec:full_sim}

Since we have all the components necessary to simulate an entire SO-SAT deep survey, we could run a limited-scale test of our complete analysis, from observing data all the way to measuring a tensor-to-scalar ratio value with a spectra-based component separation pipeline. Since we only simulate two of the SO frequency bands, f090 and f150, we use \textit{Planck} observations as ancillary data to perform component separation.\footnote{The real SO deep survey will eventually observe 6 frequency bands in total and will not need to use ancillary data from other experiments.} Since this test has a limited scale, we compromise by only simulating the 353\,GHz \textit{Planck} band. On the other hand, we do not include synchrotron as a polarised foreground in our simulation, since we do not include the low-frequency bands necessary to constrain it, but only thermal dust.

The 100 CMB realisations we use only have lensing $B$-modes and $r=0$. The thermal dust model is \texttt{d1} from \textsc{pysm}. In Section~\ref{sec:input_simulations} we described how these simulations are generated. We observe the 353\,GHz \textit{Planck} simulations with the same schedule and filtering described in Sections~\ref{sec:schedule} and \ref{sec:using_toast} with \textsc{toast}. In this paper, we work under the simplified scheme that f090 and f150 have been observed and filtered in the same way. This can be easily generalised to the point where each frequency channel is assigned its own filtering recipe. This filtering scheme is applied over an entire simulated \textit{Planck} 353\,GHz map, which includes the noise. We have 100 realisations of the \textsc{npipe} 353\,GHz frequency maps, with their A/B detector split. For the two SAT bands, we define four quarter-mission (QM) splits. The pure power-law simulations, used to estimate the transfer function, are co-added with these same splits, and the TFs are calculated for the corresponding split combinations and used to calculate the corresponding cross-spectra. Finally, we estimate unbiased spectra on the simulated observations of CMB+dust+noise by averaging all cross-spectra combinations, avoiding all auto-spectra. For example, the f090$\times$f090 spectrum is calculated from the 6 cross-spectra combinations between the 4 QM splits, the f090$\times$f150 spectrum is calculated from the 16 cross-spectra combinations, etc. For spectra, we use the full South patch mask constructed from the hits map shown in Figure~\ref{fig:hits_full}.

To estimate a covariance for error bars and parameter posteriors with Markov chain Monte Carlo (MCMC) sampling, we use 100 realisations of Gaussian dust and calculate their deconvolved spectra in the same way as described above. The total covariance is estimated from the 100 realizations of CMB + noise + Gaussian dust.

\begin{figure}
    \centering
    \includegraphics[width=1\textwidth]{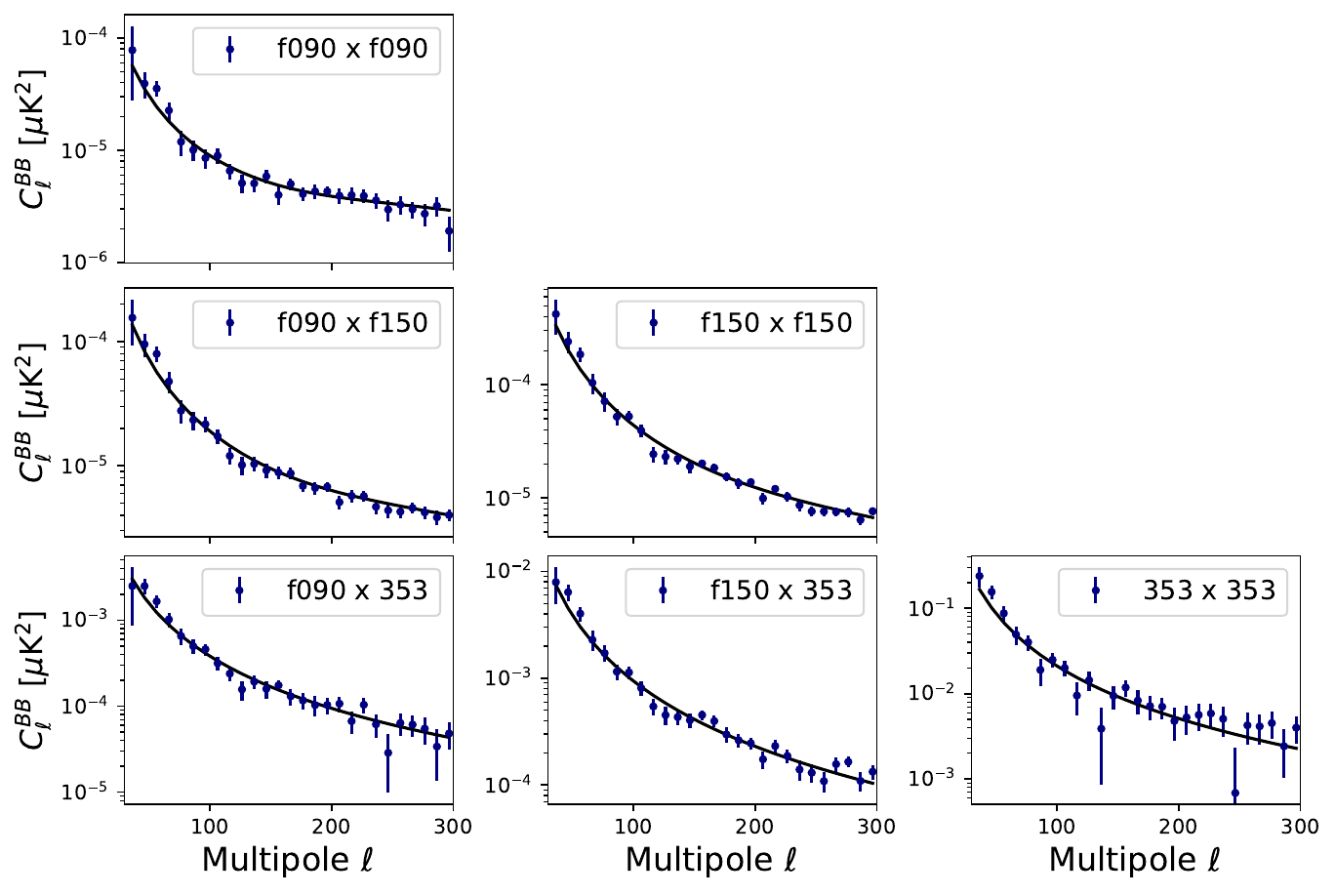}
    \caption{
    $BB$ power spectra for our limited-scale 1-year deep survey simulation. Unbiased spectra are calculated from the average of the cross-spectra of splits. The spectra for one realisation are shown in filled blue dots, while the error bars are calculated from the standard deviation across 100 realisations of the simulations that include Gaussian dust. The best-fit CMB+dust model is shown as the black solid line.
    }
    \label{fig:BB_spectra}
\end{figure}

The unbiased $C_{\ell}^{BB}$ spectra are shown in Figure~\ref{fig:BB_spectra} in filled blue dots. This is the spectrum for one realisation. The error bar is calculated from the standard deviation across 100 realisations of the simulations that include the Gaussian dust.

\begin{table}
\begin{threeparttable}
    \centering
    \caption{Parameters and priors used in our MCMC analysis. TH is a uniform top-hat prior in the given range, while G is a Gaussian prior with the mean $\pm$ the standard deviation.
    \label{table:params}
    }
    \begin{tabular}{l c c c c c}
        \hline
        \hline
        Parameter & $r$ & $A_{\rm lens}$ & $A_{\rm d}^{BB}$ & $\alpha_{\rm d}^{BB}$ & $\beta_{\rm d}$ \\
        \hline
        Units & -- & -- & $\mu$K$^2$ & -- & -- \\ 
        Prior type & TH & TH & TH & TH & G \\
        Prior value & [$-0.1,+0.1$] & [$0,2$] & [$0,60$] & [$-1.0,+0.8$] & $1.54 \pm 0.12$ \\
        \hline
        Fiducial value & 0 & 1 & $36.90 \pm 1.17^{\dagger}$ & $-0.10 \pm 0.03^{\dagger}$ & $1.55 \pm 0.01^{\dagger}$  \\
        \hline
    \end{tabular}
    \begin{tablenotes}
        \small
        \item $^{\dagger}$ These are fitted to the pure unfiltered \texttt{d1} dust model, and as such are shown here with their $1\sigma$ error bar. We used the fitted value as the fiducial value of the parameter.
    \end{tablenotes}
\end{threeparttable}
\end{table}

Our next step is to do component separation and estimate posterior probabilities for parameters, in particular $r$. We do this with the spectra-based \textsc{bbpower} pipeline,\footnote{\url{github.com/simonsobs/BBPower}} which we have used before to forecast the performance of SO~\cite{Abitbol_2021,Azzoni_2021,Azzoni_2023,2024PhRvD.110d3532H}. We choose this method for component separation, as it is the only foreground removal approach for which efficient and accurate methods exist to account for the impact of timestream- and/or map-level filtering~\citep[e.g.][]{2014PhRvL.112x1101B,2020JCAP...12..045C}. We fit a model only to the $BB$ spectrum, with five free parameters: $r$, a lensing amplitude $A_{\rm lens}$, a dust power spectrum amplitude (at pivot scale $\ell=80$ and pivot frequency $353$ GHz) $A_{\rm d}^{BB}$, a power-law index of the dust power spectrum $\alpha_{\rm d}^{BB}$, and a frequency spectral index $\beta_{\rm d}$, corresponding to the assumed MBB spectral energy distribution of Galactic dust emission. The temperature $T_{\rm d}=20$\,K of the dust MBB is fixed in our likelihood model. For each parameter, we have a fiducial value. The CMB realisations are generated with no primordial $B$-modes ($r=0$), and $A_{\rm lens}=1$, while \texttt{d1} dust has spatially variable SED spectral parameters and Equation~\eqref{eq:dust_Dell} is a good approximation within small patches on the sky, but it is not exact for a larger patch such as the South Galactic patch. We wish to determine these fiducial parameters for the \texttt{d1} dust template. We fit these parameters to the pure unfiltered \texttt{d1} template within our common hits map at the three relevant frequencies: 93, 145, and 353\,GHz, computing spectra in the multipole range $\ell=30-300$ and within our mask. As error bars for the fit, we use the standard deviation across the 100 realisations of unfiltered Gaussian dust. We use the \textsc{curve\_fit} method from \textsc{scipy}~\cite{2020NatMe..17..261V} to perform the fit. These fiducial values and priors are listed in Table~\ref{table:params} for the five parameters. While we list the dust fiducial values with their $1\sigma$ uncertainties, we use the fitted value alone.

\begin{figure}
    \centering
    \includegraphics[width=0.495\textwidth]{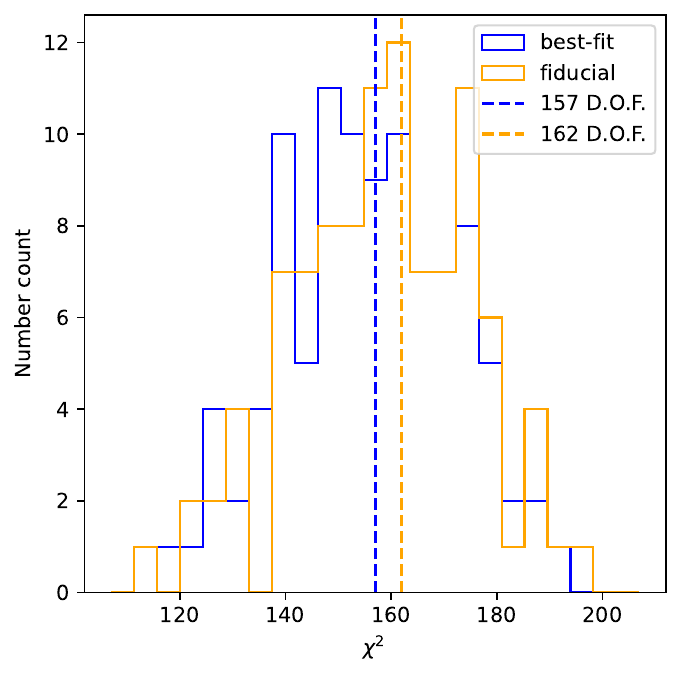}
    \includegraphics[width=0.495\textwidth]{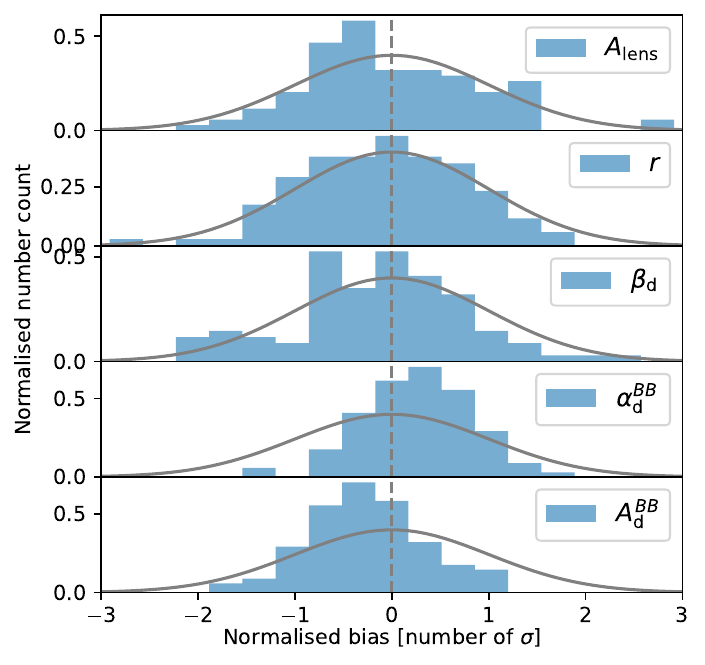}
    \caption{{\sl Left panel:} distribution of $\chi^2$ values for the $BB$ power spectra with respect to the input model (orange) and with respect to the best-fit model of each simulation (blue). The vertical lines mark the number of elements in the data vector in each case. {\sl Right panel:} distribution of the normalised bias for the best-fit parameters in each simulation. For parameter $\theta$, these are estimated as $(\theta_{\rm BF}-\theta_{\rm in})/\sigma_\theta$, where $\theta_{\rm BF}$ and $\theta_{\rm in}$ are the best-fit value and the input value for the parameter, and $\sigma_{\theta}$ is its inferred posterior standard deviation. For reference, we also show a normal distribution with unit variance $\mathcal{N}(0,1)$. Small deviations from this reference distribution are not unexpected, as the posterior distributions of these parameters are not exactly Gaussian.}
    \label{fig:chi2}
\end{figure}

We run \textsc{bbpower} implementing MCMC from \textsc{emcee}~\cite{2013PASP..125..306F} with 30 walkers and 10,000 iterations. We repeat for each of the 100 realisations. Figure~\ref{fig:chi2}, left, shows the histogram of the $\chi^2$ values for the $C_{\ell}^{BB}$ spectrum. Since we have six distinct cross-frequency $BB$ power spectra, each with 27 multipole bins between $\ell=30-300$, and five fitted parameters, the number of degrees of freedom (DOF) for our best-fit chi-squared distribution is 157, shown as the blue dashed vertical line. In blue we show the $\chi^2$ calculated with respect to best-fit values. In orange, we show the $\chi^2$ calculated with respect to fiducial values listed in Table~\ref{table:params}. The DOF in this case is 162, without fitted parameters. This is shown as the orange dashed vertical line. 

\begin{figure}
    \centering
    \includegraphics[width=0.85\textwidth]{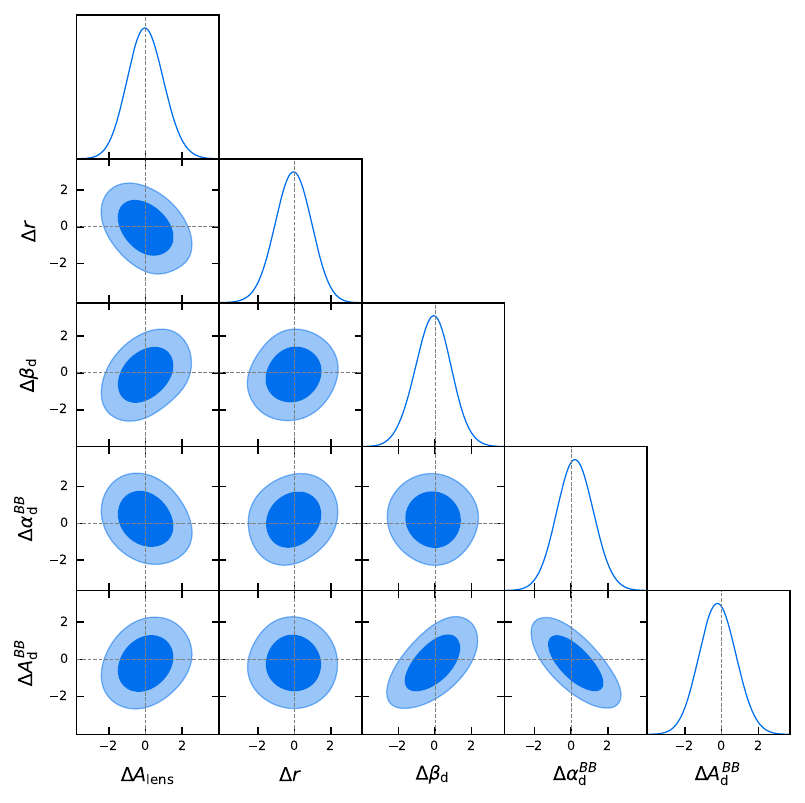}
    \caption{
    Triangle plot of the posterior distributions of residual parameters for all 100 realisations. We concatenated the chains for the 100 realisations and estimated the posteriors over that. For each parameter value, we subtract the fiducial value listed in Table~\ref{table:params} and divide by the $1\sigma$ uncertainty.
    }
    \label{fig:triangle_plot}
\end{figure}

In order to focus on the ability of our pipeline to obtain unbiased constraints from the first year of SO-SAT data, and not on the forecasted sensitivity, we present our simulated cosmological constraints normalised by their standard deviation, referred to in this work as the normalised bias.\footnote{We subtract the fiducial value from the parameter and divide by the $1\sigma$ error bar.} 
The histogram for the normalised bias for the best-fit value for each parameter is shown in Figure~\ref{fig:chi2}, right. Figure~\ref{fig:triangle_plot} shows the posterior distributions for our five parameters for the 100 realisations. For each parameter value, we subtract the fiducial value and divide by the $1\sigma$ uncertainty.
\section{Summary and conclusions} \label{sec:conclusions}

In this work, we described methods to estimate pseudo-$C_\ell$ power spectra from biased filter+bin maps in the context of small aperture telescope surveys in CMB experiments, looking for the signature of primordial gravitational waves. The maps are biased due to timestream filtering and there is no analytical formula to account for the mode coupling or polarisation leakage induced on the pseudo-$C_\ell$ estimate. Hence, simulations are required, where the exact mixing can be estimated by measuring how individual $\ell$s mix into one another. However, this approach is limited by the cosmic variance of the maps and therefore a huge number of realisations ($\mathcal{O}(10^4)$) is required to reduce the estimator noise down to a practical level. A much computationally cheaper alternative approach is to assume the mixing is the analytical solution from MASTER for a masked sky, suppressed by a transfer function measured in multipole band powers, under the approximations described in Section~\ref{sec:method_tf}. This transfer function can be cheaply estimated averaging over $\mathcal{O}(100)$ filtered realisations of pure $E/B$ modes.

The main objective of this work is to validate this method with simulations of the deep survey of the Simons Observatory. Using the \textsc{toast} software, we simulate the observing of the CMB during a one-year season, scanning into TODs and simulating the modulation by a rotating HWP. We also demodulate and apply time-domain filtering to produce filter+bin maps. However, for the real SO data, we do not expect to use filtering as aggressive as what we use in this work. Besides scanning pure signal maps, we also simulate a realistic noise with the same scanning schedule.

While the transfer function estimator can be used at smaller scales, it is biased by $>1 \sigma$ in the large-scale limit $\ell \sim 30$. We illustrate this by performing a simple null test of rising versus setting scans. Besides CMB, we include foreground emission and noise consistent with a one-year deep survey with two SO-SATs. In this case, the bias induced in the largest-scale band powers is small compared to the statistical noise introduced, so such a null test is passed.

Finally, we perform an end-to-end analysis, starting from scanning the sky, through making maps, calculating the cross-spectra, and doing likelihood fitting cosmological and foreground parameters. By calculating a $\chi^2$ statistic, as exemplified in Figure.~\ref{fig:chi2}, we show that our spectra are unbiased. We also show that the constraints on the fitted parameters (especially $r$ and $A_{\rm lens}$) are unbiased. We conclude that the transfer function method described in this paper is validated for a survey with a noise level $\sim 5$\,$\mu$K-arcmin at 90\,GHz, and thus is valid for use in the analysis of the first 1-2 years of SO data.

\acknowledgments
CHC acknowledges ANID FONDECYT Postdoc Fellowship 3220255 and BASAL CATA  FB210003. 
KW is supported by the STFC, grant ST/X006344/1, and by a Gianturco Junior Research Fellowship of Linacre College, Oxford. 
ALP and DA acknowledge support from an STFC Consolidated Grant (ST/W000903/1). 
CB acknowledges partial support by the Italian Space Agency LiteBIRD Project (ASI Grants No. 2020-9-HH.0 and 2016-24-H.1-2018), as well as the InDark and LiteBIRD Initiative of the National Institute for Nuclear Physics, and the RadioForegroundsPlus Project HORIZON-CL4-2023-SPACE-01, GA 101135036. 
MLB and DBT acknowledge support from the UKRI/STFC (grant number ST/X006344/1). 
EC acknowledges support from the Horizon 2020 ERC Starting Grant (Grant agreement No 849169).
YC acknowledges the support from JSPS KAKENHI Grant Number JP24K00667.
RD thanks ANID for grants BASAL CATA FB210003, FONDEF ID21I10236 and QUIMAL240004. 
JE, SB, ETKS, WS,AVA acknowledge the SCIPOL project funded by the ERC under the European Union’s Horizon 2020 research and innovation program (Grant agreement No. 101044073). 
GF is supported by the STFC Ernest Rutherford fellowship. 
SG acknowledges support from STFC and UKRI (grant numbers ST/W002892/1 and ST/X006360/1). 
MM is funded by the European Union (ERC, RELiCS, project number 101116027). Views and opinions expressed are however those of the author(s) only and do not necessarily reflect those of the European Union or the European Research Council Executive Agency. Neither the European Union nor the granting authority can be held responsible for them.

The Geryon cluster at the Centro de Astro-Ingenieria UC was extensively used for the calculations performed in this paper. ANID BASAL project FB21000, BASAL CATA PFB-06, the Anillo ACT-86, FONDEQUIP AIC-57, and QUIMAL 130008 provided funding for several improvements to the Geryon cluster. 
This research used resources of the National Energy Research Scientific Computing Center (NERSC), a Department of Energy Office of Science User Facility using NERSC award HEP-ERCAP-mp107.
This work was supported in part by a grant from the Simons Foundation (Award \#457687, B.K.).
This research has made extensive use of \textsc{numpy}~\cite{2020Natur.585..357H}, \textsc{scipy}~\cite{2020NatMe..17..261V}, \textsc{matplotlib}~\cite{2007CSE.....9...90H}, \nmt~\cite{2019MNRAS.484.4127A}, \textsc{healpy}~\cite{2019JOSS....4.1298Z}, \textsc{emcee}~\cite{2013PASP..125..306F}, \textsc{pysm}~\cite{2017MNRAS.469.2821T}, and \textsc{getdist}~\cite{2019arXiv191013970L}.

\appendix
\section{Effect of individual filters} \label{sec:individual_filters}

\begin{figure}
    \centering
    \includegraphics[width=0.85\textwidth]{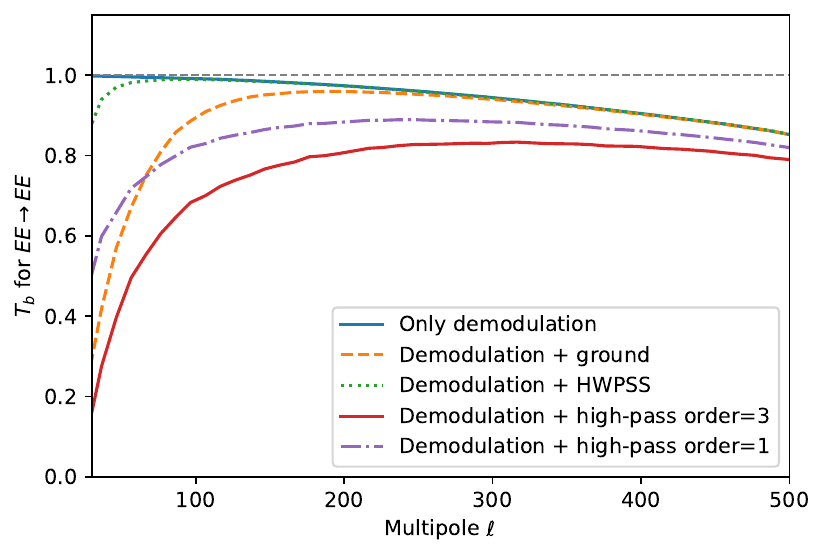}
    \caption{TF estimated for individual filters. The blue solid line shows the effect of only demodulating the HWP, while other lines show the effect of demodulation plus another filter. This is estimated from the average of 100 realizations of power-law simulations and on a fraction of the full schedule used in this paper. This is for the f090 frequency channel.
    }
    \label{fig:TF_plateau}
\end{figure}

Here, we describe the features of the transfer function presented in Section~\ref{sec:tf_full_multipole} and Figure~\ref{fig:tf_cmb}. Figure~\ref{fig:TF_plateau} shows transfer functions estimated for different combinations of individual filters. This is calculated from 100 realizations of power-law simulations and from a fraction of the full schedule described in Section~\ref{sec:schedule}. 

The first filter applied is the demodulation of the HWP, which has no effect on large scales and a minimal effect towards smaller scales, since we low-pass the timestreams at 2\,Hz after demodulating, so higher frequencies (and therefore smaller scales) are completely suppressed.\footnote{Note that there is some interplay between the resolution of the pixelated input map and the low-pass filter of the demodulation, where there might be a fraction of power above $\sim 2$\,Hz that is lost when low-pass filtering. This effect is more pronounced at $N_{\rm side}=512$,  the resolution of our simulations, than at a higher $N_{\rm side}$ like 2048, and therefore all of the transfer functions in this plot would show an extra suppression towards higher $\ell$.} The HWP synchronous filter has effect on very large scales, while the ground filter induces a very large suppression of the very large-scale modes. Overall, demodulation, HWP synchronous signal and ground filters do not show suppression on smaller scales $\ell \sim 300$ and therefore the TF should approach $T_b \sim 1$.

On the other hand, the full TF shown in Figure~\ref{fig:tf_cmb} plateaus at a lower value $T_b \sim 0.8$. As we can see here, the high-pass polynomial filter is responsible for this (shown in the solid red and dashed-dotted purple lines), which suppresses the large scales by a large amount, and suppresses the middle-to-small scales by $\sim 20$\%. Moreover, a polynomial of higher order 3 suppresses more modes than a polynomial of order 1. In this paper, we use a high-pass filter with order 3. To analyse real data, a pipeline with these filtering choices adjusted is already being used to reduce the impact of these transfer functions.

\section{Illustration of focal plane thinning} \label{sec:fp_thinning}

\begin{figure}
    \centering
    \includegraphics[width=1.0\textwidth]{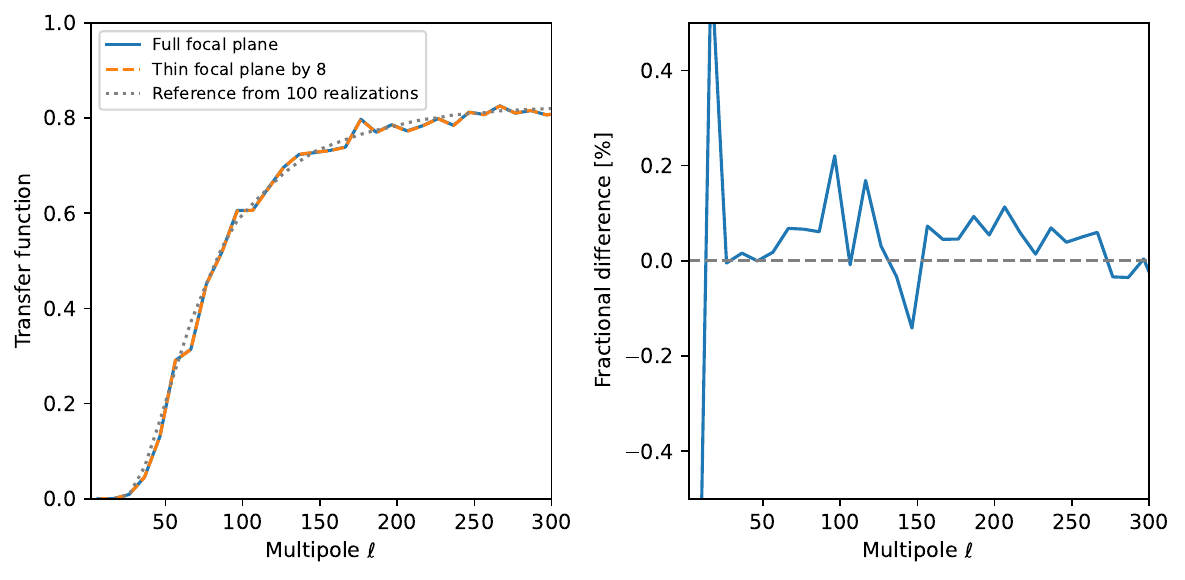}
    \caption{Illustration of the effect of focal plane thinning. $EE$ TF estimated with a single simulated power-law map, using both the full focal plane and thinning it by a factor of 8. We also show the smooth reference TF estimated from 100 realizations in dotted grey (the same shown in Figure~\ref{fig:tf_cmb}). The fractional difference, as shown in the right panel, is within 0.2\% for most of the multipole range.}
    \label{fig:compare_thinfp}
\end{figure}

To ease the computation of simulations in \textsc{toast}, we thin the focal plane by 8, i.e., we only simulate one for every eight detectors. Figure~\ref{fig:compare_thinfp} shows the $EE$ TF calculated from a single power-law simulated map, which is filtered both using the full focal plane as well as thinning the focal plane by a factor of 8. The fractional difference is ${\sim} 0.1$\% for most of the multipole range, increasing to ${\sim} 1$\% only for the first couple of bins at $\ell < 30$.

\bibliography{biblio}
\end{document}